\documentclass[amsmath, reprint, floatfix, superscriptaddress, aps, prx]{revtex4-2}
\usepackage{booktabs}
\usepackage{upgreek}
\usepackage{orcidlink} 

\begin{document}
\title{Remote sensing of tectonic induced stress across faults using high energy muon beams}

\author         {L. Serafini\,\orcidlink{0000-0002-4367-1555}}
\affiliation    {INFN-Milano, Via Celoria 16, 20133, Milan, Italy}

\author         {G. Muttoni\,\orcidlink{0000-0001-7908-1664}}
\affiliation    {University of Milan, Dept. Earth Sciences DISTAD, Via Mangiagalli 34, 20133, Milan, Italy}

\author         {A. Bacci\,\orcidlink{0000-0001-6010-9225}} 
\affiliation    {INFN-Milano, Via Celoria 16, 20133, Milan, Italy}

\author         {F. Broggi\,\orcidlink{0000-0003-1628-1604}}
\affiliation    {INFN-Milano, Via Celoria 16, 20133, Milan, Italy}

\author         {L. Giuliano\,\orcidlink{0000-0002-5326-1971}}
\affiliation    {INFN-Roma 1, Piazzale Aldo Moro 2, 00185, Rome, Italy}
\affiliation    {SBAI Department, Sapienza University of Rome, 00161 Rome, Italy}

\author         {A. M. Marotta\,\orcidlink{0000-0002-9580-5981}}
\affiliation    {University of Milan, Dept. Earth Sciences DISTAD, Via Mangiagalli 34, 20133, Milan, Italy}

\author         {V. Petrillo\,\orcidlink{0000-0002-8556-3384}}
\affiliation    {INFN-Milano, Via Celoria 16, 20133, Milan, Italy}
\affiliation    {University of Milan, Dept. of Physics, Via Celoria 16, 20133, Milan, Italy}

\author         {E. Puppin\,\orcidlink{0000-0002-8382-3621}}
\affiliation    {INFN-Milano, Via Celoria 16, 20133, Milan, Italy}
\affiliation    {Politecnico di Milano, Dept. of Physics, P.za Leonardo Da Vinci 32 , 20133, Milan, Italy}

\author         {M. Rossetti Conti,\orcidlink{0000-0002-5767-3850}}
\email{marcello.rossetti@mi.infn.it}
\affiliation    {INFN-Milano, Via Celoria 16, 20133, Milan, Italy}

\author         {A. R. Rossi\,\orcidlink{0000-0002-6216-8664}}
\affiliation    {INFN-Milano, Via Celoria 16, 20133, Milan, Italy}

\author         {S. Samsam\,\orcidlink{0000-0001-6311-3801}}
\affiliation    {INFN-Milano, Via Celoria 16, 20133, Milan, Italy}

\author         {M. Voltolini\,\orcidlink{0000-0002-3843-4854}}
\affiliation    {University of Milan, Dept. Earth Sciences DISTAD, Via Mangiagalli 34, 20133, Milan, Italy} 

\author         {M. Zucali\,\orcidlink{0000-0003-3600-7856}}
\affiliation    {University of Milan, Dept. Earth Sciences DISTAD, Via Mangiagalli 34, 20133, Milan, Italy}

\date{\today}
\begin{abstract}
We illustrate a theoretical study of a newly conceived technique using high-energy muon beams (TeV-class) propagating through thick (km-long) crystalline rock layers subject to tectonic-induced stress, potentially capable of actively monitoring the temporal evolution of the pressure rise in seismic fault zones associated with earthquake triggering when the induced tectonic pressure reaches and overcomes the rock elasto-plastic deformation limit.
This technique could contribute to improving earthquake forecasting statistics in seismically active regions, offering support for seismic hazard assessment and prevention strategies.

Active monitoring of the induced tectonic stress and its time evolution is achieved by remote sensing of the electric field generated in quartz crystals embedded in crystalline rocks by piezoelectric effects. In this context, tectonic pressure refers to the time-dependent stress field acting on the rock body due to tectonic forces, which adds to the time-independent lithostatic pressure resulting from the weight of overlying materials.
High-energy muon beams transmitted through a rock layer subject to tectonic pressure will be affected in their transverse phase space distributions by the piezoelectric fields, therefore transferring to a detector the information on the applied tectonic stress.

Finally, we illustrate the design of a proof-of-principle experiment to be conducted in a standard accelerator laboratory, using moderate-energy muons (GeV-class) propagating through granite slabs subject to a press-induced stress reaching the rupture limit.
A zero-generation proof-of-principle test can also be performed using 20-150\,MeV electron beams transmitted through single quartz crystals subject to variable pressure.
\end{abstract}

\maketitle

\section{Introduction} \label{sec:Intro}
The global annual toll of casualties resulting from earthquakes underscores the crucial importance of advancing seismic risk mitigation strategies.
Seismic risk prevention relies on the integration of three core elements: the assessment of seismic hazard through the identification and probabilistic modeling of active fault systems; the reduction of structural vulnerability via earthquake-resistant design and retrofitting measures; and the preparation of emergency response plans, including public education and coordination of civil protection strategies. 
Together, these components aim to reduce both structural damage and human casualties in seismically active regions, but experience shows that their implementation remains largely limited to developed countries.

Improving the assessment of seismic hazards through innovative approaches such as the technique proposed in this study is essential to strengthen the forecasting component of earthquake prevention.

In this paper, we describe a possible technique capable of continuously monitoring the tectonic pressure building up across fault zones, which leads to the exceedance of the rock's elasto-plastic rupture limit and ultimately triggers earthquake events.
Maximum depths up to about 10\,km can be investigated, a region of the crust characterized by frequent destructive earthquakes.

\begin{figure}[htbp]
    \centering
    \includegraphics[width=.95\linewidth]{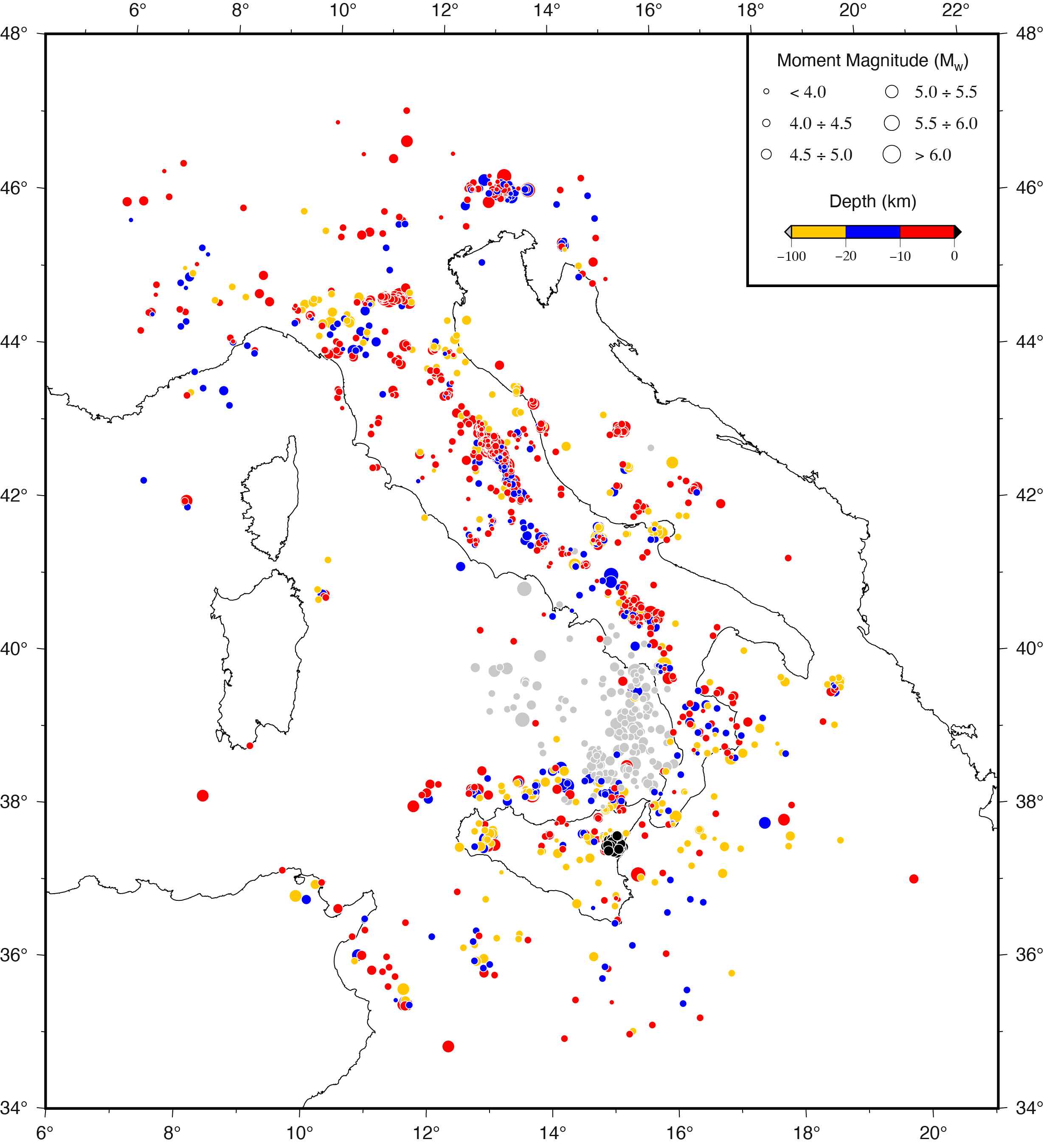}
    \caption{Map of the Italian seismicity from the CPTI15 v.4 Parametric Catalog of the Italian Earthquakes \cite{RovidaEtAl2022}, as function of depth and Moment Magnitude.
    Only earthquakes with known depth and Moment Magnitude are included.
    Earthquakes with depths greater than 100\,km are shown in grey.}
    \label{fig:seismicity_map}
\end{figure}

\begin{figure}[htbp]
    \centering
    \includegraphics[width=.9\linewidth]{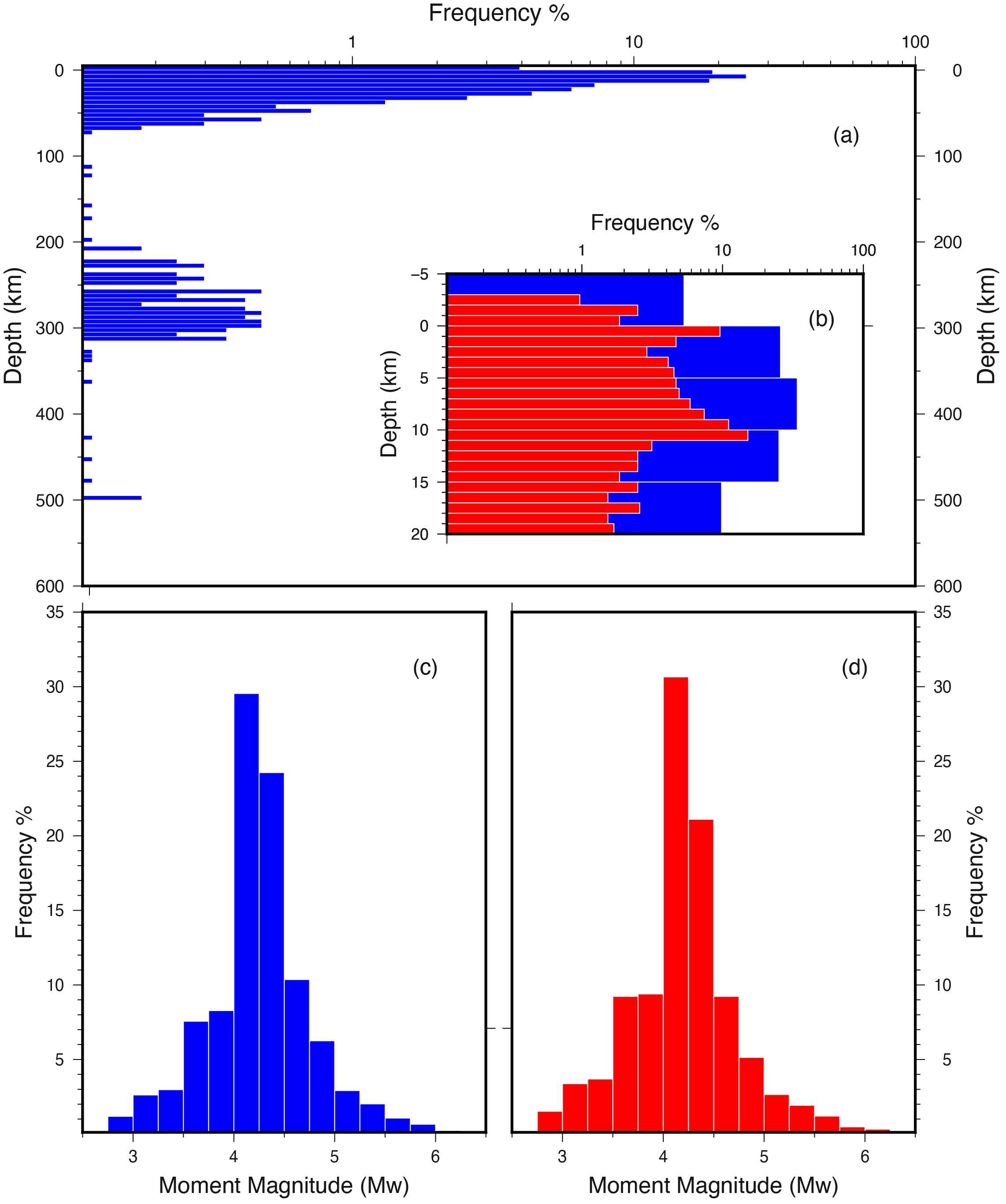}
    \caption{(a) Earthquake depth distribution for all events shown in \autoref{fig:seismicity_map}.
    A bin interval of 5\,km is used.
    Two main depth clusters are evident, from 0 to about -80\,km depth and from -200\,km to -350\,km.
    (b) Detail of the earthquake depth distribution shown in panel (a) for events with depths shallower than -20\,km.
    Bins of 5\,km (blue color) and 1\,km (red color) are used.
    (c) Earthquake Moment Magnitude distribution for the same events as in panel (a).
    A bin of 0.25 is used.
    (d) The same as in panel (c) but only for earthquakes with depths shallower than -20\,km.}
    \label{fig:depth_magnitude_distributions}
\end{figure}

Inspection of the distribution map of earthquakes recorded in Italy from 1918 to 2020 (\autoref{fig:seismicity_map}), along with diagrams showing the distribution of occurrences as a function of depth and moment magnitude (\autoref{fig:depth_magnitude_distributions}), reveals that approximately 55$\%$ of the events occurred at depths less than 10\,km (74$\%$ at depths less than 20\,km) and, among these, approximately 70$\%$ had a magnitude moment equal to or greater than 4 (73$\%$ for events with depths less than 20).

    
The technique is applicable mainly to quartz-rich crystalline rocks, since it exploits the piezoelectric field generated in quartz crystals, which induces a transverse deflection on the charged particle propagating through the crystal itself.
Charged particles routinely employed in medical, industrial and research applications, such as electrons, protons and ions, or neutral particles, such as neutrons or photons, are not capable of penetrating rock layers on the km-thick scale.
Only muons with energies larger than hundreds of GeV can survive such a journey. 
Neutrinos have also such capability, even higher in terms of penetrability, but they are not easy to produce and detect (and they are not, or very weakly, affected by electro-magnetic fields). Furthermore, muon beams in this range of energy, produced by particle accelerators, are fully consistent with the objectives of muon colliders, which are nowadays intensively under development by the high-energy physics community.

\begin{figure}[htbp]
    \centering
    \includegraphics[width=1\linewidth]{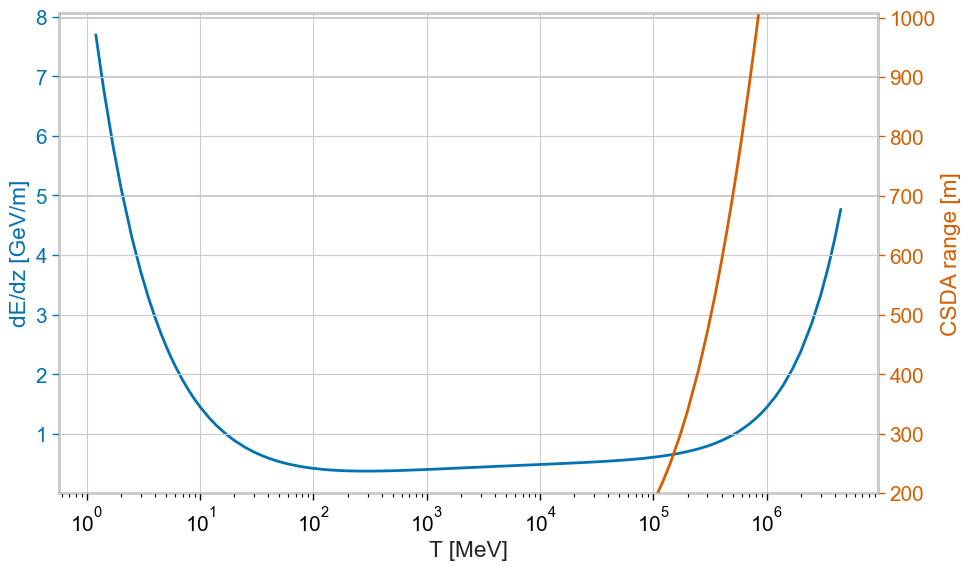}
    \caption{Differential energy loss ($dE/dz$) of muons, measured in GeV per meter of propagation in silicon dioxide (fused quartz, SiO$_2$), as a function of the muon kinetic energy $T$ (in MeV, on a logarithmic scale). The left $y$-axis shows $dE/dz$, while the right $y$-axis displays the CSDA (Continuous Slowing Down Approximation) range in meters, which represents the mean path length a muon travels before stopping. Both quantities are extracted from \cite{ParticleDataGroup:2024cfk} data.
    }
    \label{fig:dE}
\end{figure}

Using high-energy TeV-class muon beams, several kilometers of rock thickness can be sampled, exploiting the large penetrability of muons in solid matter.
\autoref{fig:dE} shows how muons with kinetic energy in the range 500\,GeV to 1\,TeV can propagate in silicon dioxide (quartz) up to approximately one kilometer of thickness. 
In principle, such high-energy muons are capable of propagating through thick layers of rock, while sampling the electric field present inside the quartz crystals, which applies mainly transverse deflections to the muons. 
The case of a real crystalline rock will be discussed later. 
The direction of the piezoelectric field is randomly distributed in space. The muons, during propagation, acquire random momenta, following a random-walk. This process is similar to another typical and well-known diffusion: the Multiple Coulomb Scattering (MCS).
However, while MCS is independent of the applied pressure, the piezoelectric induced random-walk diffusion is dependent on the pressure time evolution and directly generates an irreversible beam emittance growth during the propagation through the rock layer across a tectonic fault.

\vspace{1 mm}

By measuring the emittance degradation of the muon beam at the end of its propagation path, it becomes possible to reconstruct and continuously monitor the effective tectonic pressure building up across the fault zone, enabling the development of an effective earthquake forecasting system.
The sensitivity of this technique to tectonic pressures is analyzed through numerical simulations, showing the potential to actively detect tectonic pressures down to a fraction of GPa at a few km depth/penetration thicknesses.
A conceptual layout of a system based on a TeV-class muon accelerator, located on the ground in a region rich of active seismic faults, can be conceived.
This scheme adopts and adapts the expected performances predicted by the vigorous R\&D program currently underway in the high-energy physics community for the development of future muon colliders \cite{collider}, which aims to accelerate, transport and control intense muon beams up to multi-TeV energies.

\begin{figure}[htbp]
    \centering
    \includegraphics[width=1\linewidth]{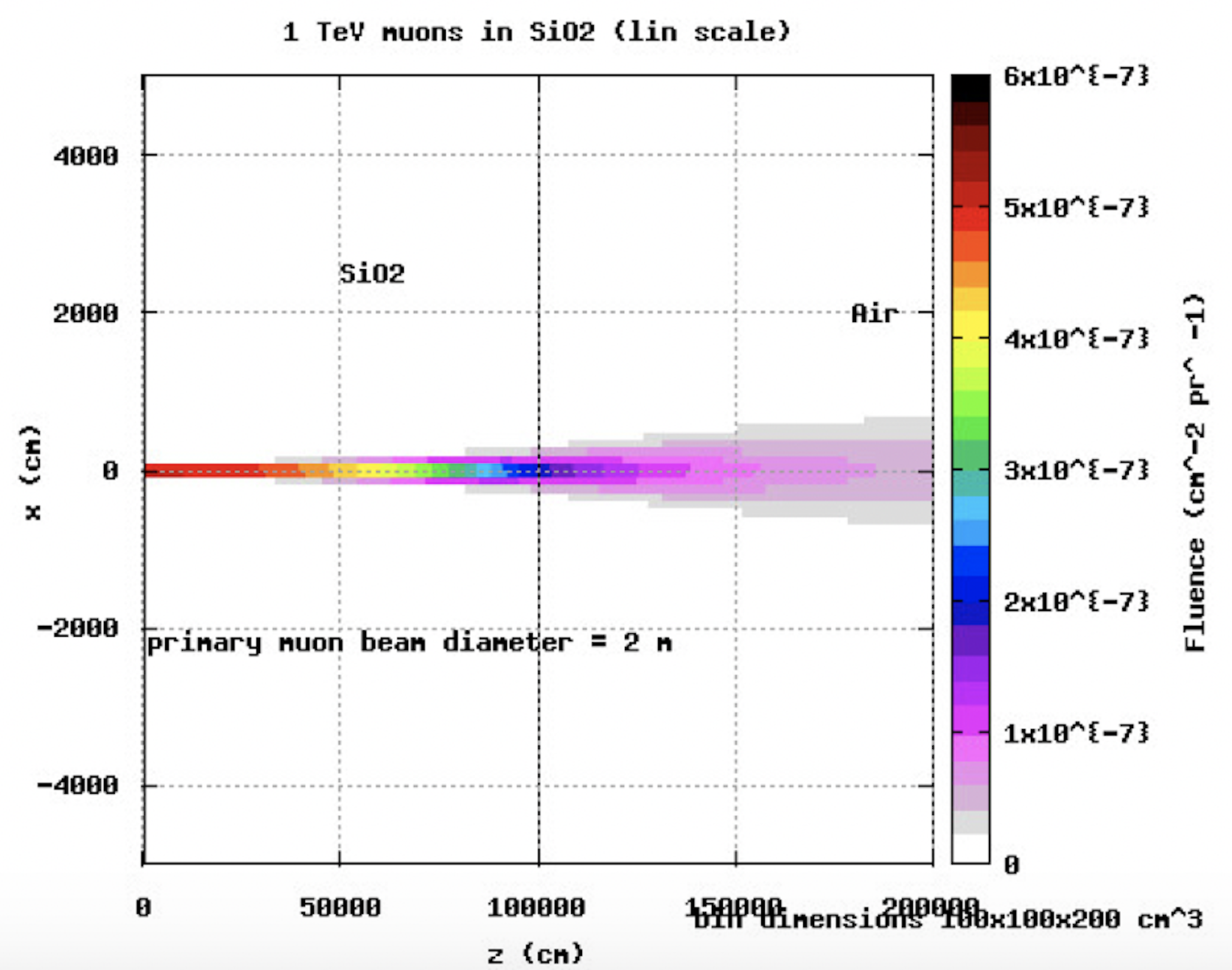}
    \caption{1\,TeV muon beam propagating into 1\,km thick pure SiO$_2$ as simulated by FLUKA: beam fluence in false colors with muon beam transverse coordinate x on the vertical axis (in\,cm)}
    \label{fig:Envelope}
\end{figure}

\begin{figure}[ht]
    \centering
    \includegraphics[width=1\linewidth]{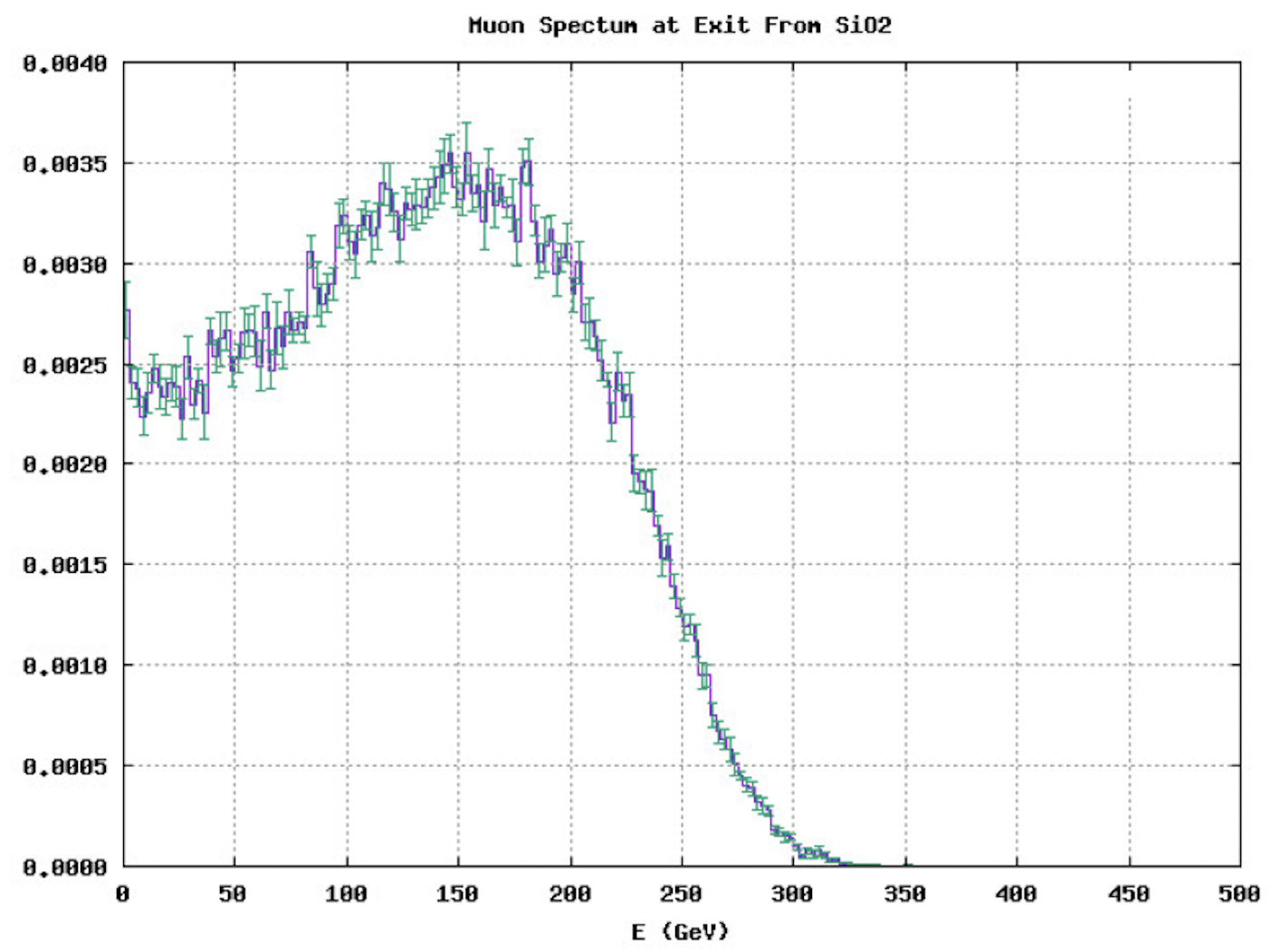}
    \caption{Muon beam energy spectrum exiting 1\,km thick SiO$_2$}
    \label{fig:Spectrum}
\end{figure}

\autoref{fig:Envelope} illustrates, as outputs from a FLUKA simulation \cite{Ahdida2022FLUKA}, the propagation of a muon beam with an initial 1\,TeV energy through 1\,km of pure SiO$_2$, followed by 1\,km in air.
The muon beam fluence is represented in false colors, demonstrating how the beam remains well collimated throughout the entire km-thick matter traversal, with a fluence decrease of about a factor 2, implying a beam diameter increase by just a factor $\sqrt{2}$ from the initial 2 meter at injection into the SiO$_2$ layer.
The distribution of the residual energy spectrum of the muons after traversing the rock is shown in \autoref{fig:Spectrum}: most muons have a residual energy greater than 50\,GeV, which makes them very adequate for a measurement of all beam characteristics and its rms momenta after the beam emerges from the rock. 
72.5 $\%$ of the initial muons survive the propagation through the 1\,km thick rock layer.

The paper is therefore organized as follows: In \autoref{sec:CrystalsUnderPressure} we discuss how the two types of pressure acting inside the Earth's crust handle the equilibrium state of crystalline rocks and how they may act as precursors of seismic events. \autoref{Sec:Theory} illustrates a theoretical model which derives the fundamental scaling laws and approximate analytical formulas predicting the muon beam spot size and divergence angle at the exit of the rock layer traversal, taking into account the combined effects of MCS and piezoelectric induced random walk.
In \autoref{sec:MuAEGIS} we illustrate our calculations, both via semi-analytical models (performed with MuAEGIS, Muon Underground Active Earthquake Genesis Investigation Software), and through detailed numerical simulations (with FLUKA), of high-energy muon beams propagating through crystalline rocks enriched in quartz crystals acting as electrostatic diodes. The results illustrate the expected perturbations in the 4D transverse phase-space distribution of the muon beam induced by lithostatic pressure within the fault zone.
\autoref{Sec:Layout} is focused on a straw-man design of a possible accelerator complex to generate the requested muon beams, based on advanced accelerator concepts and modern muon beam transport techniques.
We conclude with \autoref{Sec:PoP} by discussing possible proof-of-principle experiments necessary to validate the theoretical predictions and the numerical simulation results.
In the conclusions, we advocate for immediate efforts from the relevant scientific communities (earth sciences, seismology, accelerator physics, high-energy physics) to invest the necessary human and instrumental resources to assess the feasibility and future implementation of the proposed technique, ultimately aiming to reduce the human impact of major seismic events.

\section{Lithostatic and Tectonic pressures in fault areas: the case with crystalline rocks}   
\label{sec:CrystalsUnderPressure}

\begin{figure}[ht]
    \centering
    \includegraphics[width=1\linewidth]{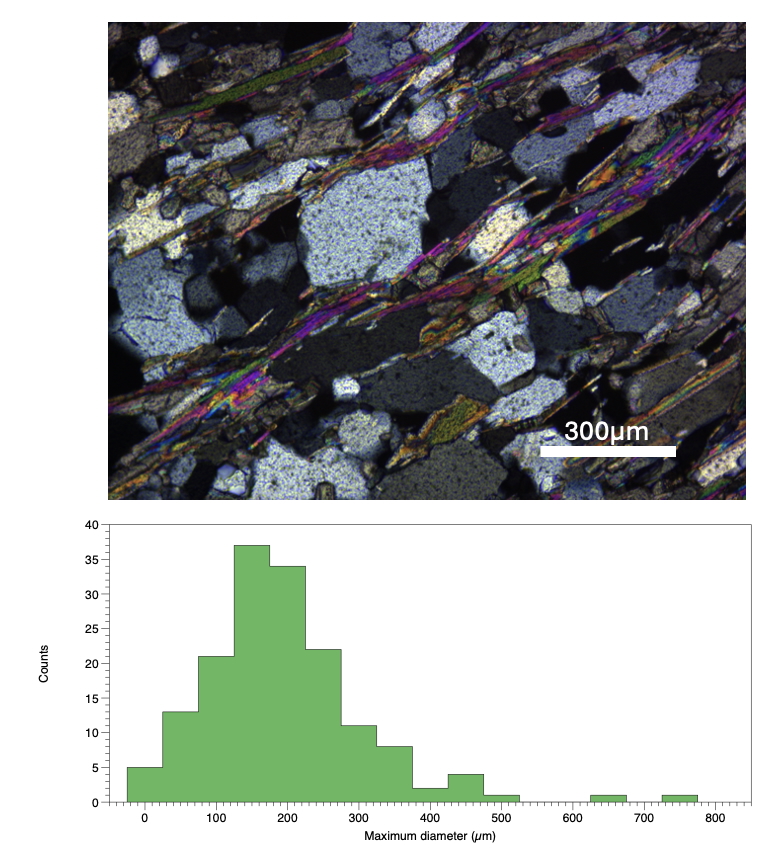}
    \caption{Optical microscopy image of a quartz-rich gneiss. Large gray individuals of quartz are bounded by colorful elongated mica grains. Histogram showing the grain-size distribution for the quartz-rich gneiss}
    \label{fig:Micro}
\end{figure}

Quartz-rich rocks such as granitoids, gneisses or sandstones constitute a large part of the continental crust. They are composed of quartz, feldspar and mica, and, depending on the tectonic evolution, other silicate minerals \autoref{fig:Micro}. In an average granitic continental crust composition, quartz may constitute up to 30\% of the crust. The tectonic evolution may also control the size and distribution of minerals within rocks, ranging from randomly distributed mm- to cm-thick poly-crystalline aggregates, to strongly aligned $\upmu$m-sized aggregates. The preferred orientation of minerals may arise from either their shape or their crystallographic structure. The latter better controls the physical properties, such as piezoelectricity, and may lead to strong coherent piezoelectric effects, as discussed in \cite{bishop1981piezoelectric}. \autoref{fig:Micro} shows an example of a polycrystalline quartz aggregate in a foliated gneiss. Sub-grains size distribution is shown in the histogram. 

\begin{figure}[htbp]
    \centering
    \includegraphics[width=.9\linewidth]{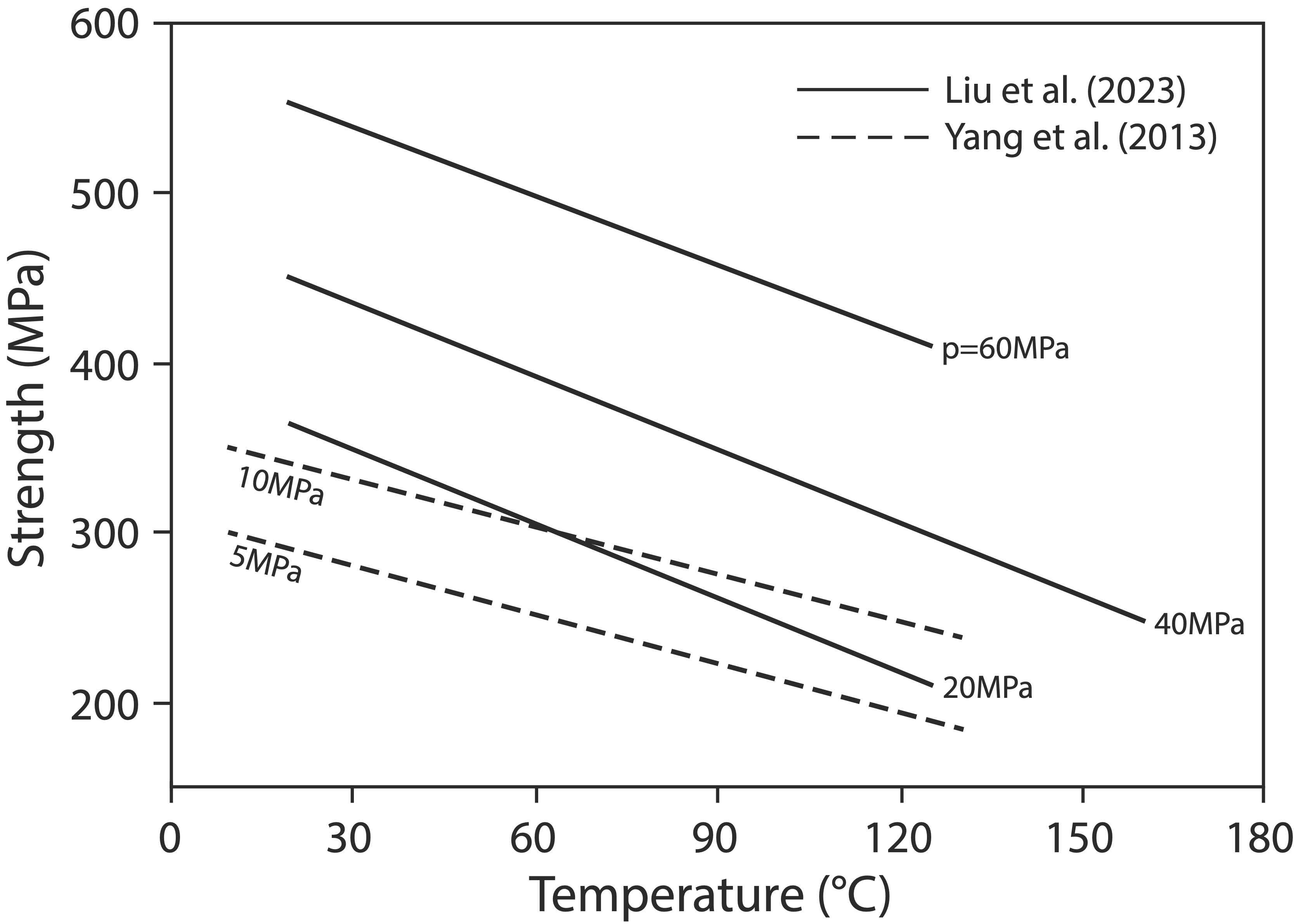}
    \caption{Empirical values of compresive strength of granite as function of temperature, for different values of confining pressure $P$. (Based on \cite{LiuEtAl2023}, continuum lines, and \cite{YangEtAl2013}, dashed lines}
    \label{fig:triaxial}
\end{figure}

In this study, we are interested in the magnitude of stress in fault areas, which can induce crustal rupture causing in turn the release of energy in the form of seismic waves associated with earthquakes.
Rock rupture occurs when the stress exceeds a limiting value, the brittle yield strength, which is  measured experimentally through typical triaxial experiments in which the strength and the way in which the failure occurs is measured under different conditions of confining pressure and temperature (e.g. \cite{Byerlee1967,Byerlee1968,Karato2008,Ramsay1967,Ranalli1995}, and references therein).
In particular, laboratory experiments indicate that the strength decreases with temperature and increases with confining pressure. \autoref{fig:triaxial} shows the results of a typical triaxial test for quartzite.

When rock failure occurs through a brittle behavior, micro-cracks develop that interact to form a fault along which the sliding occurs.
Since most rock systems have preexisting micro-cracks, their strength is lower than that of an intact rock \cite{PatersonWong2005,Karato2008} and it is controlled by the resistance against sliding. One of the most widely used failure criterion is the Mohr--Coulomb criterion that states that sliding occurs when the magnitude of the shear stress $\boldsymbol{\tau}$ exceeds a critical value $\tau_s$ defined as:
\begin{equation}
    |\boldsymbol{\tau}| \ge \tau_s = \tau_0 + f \sigma_n,
    \label{eq:mohr-coulomb}
\end{equation}
where $\sigma_n$ is the normal stress, $\tau_0$ is the cohesive strength (the resistance in absence of normal stress) and $f$ is the coefficient of internal friction that, for most rocks, varies in the range $0.5 - 0.8$ ($0.6 \le f \le 0.7$ for quartzite) (\autoref{fig:coulomb}).

\begin{figure*}[htbp]
    \centering
    \includegraphics[width=.9\linewidth]{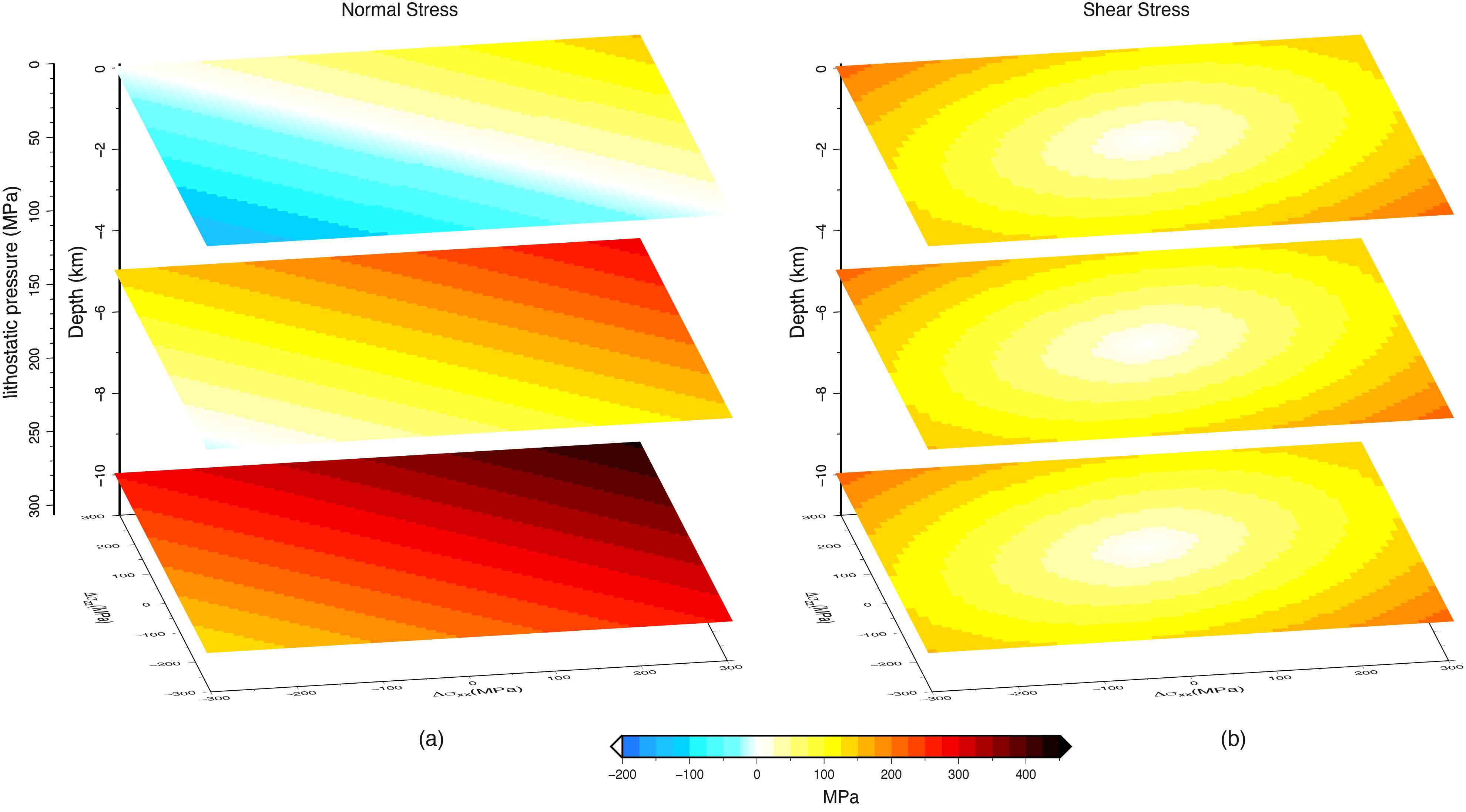}
    \caption{Values of the normal (panel a) and shear stress (panel b), calculated on a fault plane with a dip of $45^\circ$ and a strike of $45^\circ$, shown at depths of 0\,km, -5\,km and -10\,km.
    A density of 2850\,kg/m$^3$ is assumed to calculate the lithostatic pressure.}
    \label{fig:stresses}
\end{figure*}

\begin{figure}[htbp]
    \centering
    \includegraphics[width=.9\linewidth]{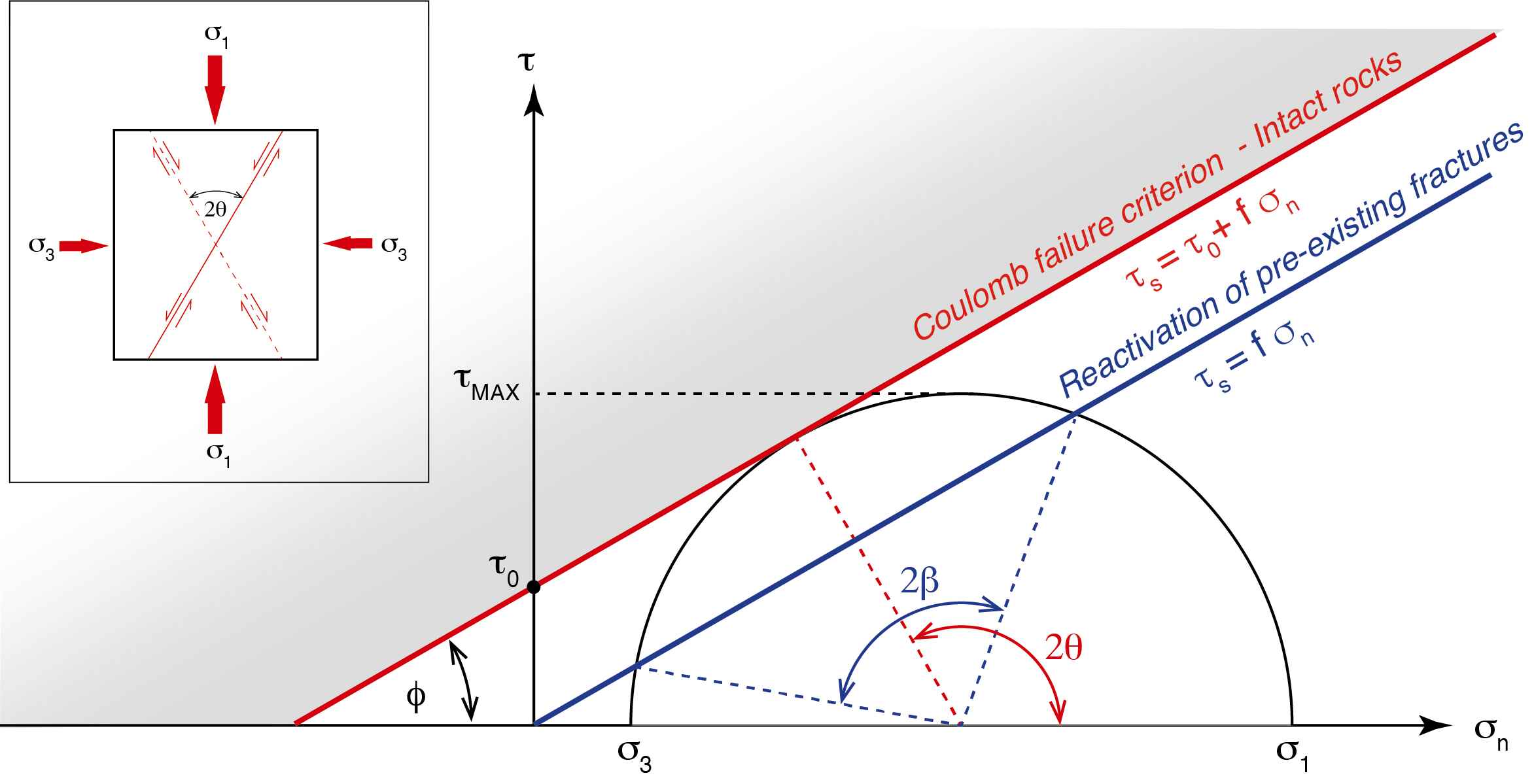}
    \caption{2D simplified graphical representation of the Coulomb failure criterion with the Mohr stress circle at failure.
    The Coulomb criterion is shown for both fracturing of intact rocks (red color), on planes with orientation $\theta$, and reactivation of pre-existing faults, having orientation comprised in the $2\beta$ angles (blue color).
    $\sigma_1$ and $\sigma_3$ are the maximum and the minimum principal stresses; $\tau_{max}$ is the maximum shear stress.
    $f=\tan{\phi}$ is the coefficient of internal friction; $\phi$ is the internal friction angle; $\tau_0$ is the cohesive strength, that is the resistance in absence of normal stress, and is equal to zero when sliding occurs along a preexisting fault.
    Inset: Geometrical configuration of the conjugate fault planes that develop when failure occurs in an intact rock.}
    \label{fig:coulomb}
\end{figure}

\begin{figure}[htbp]
    \centering
    \includegraphics[width=.9\linewidth]{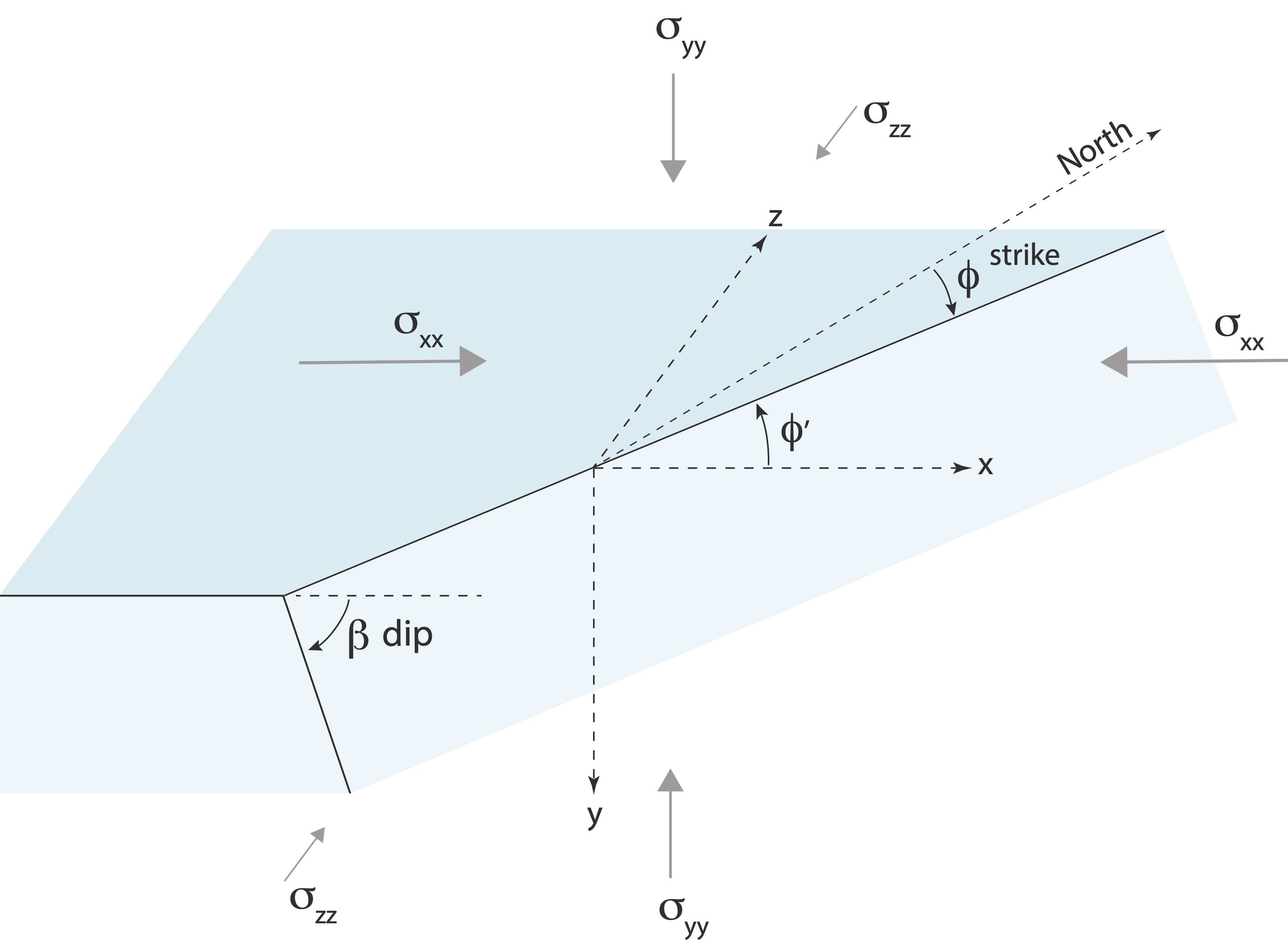}
    \caption{Scheme used to calculate the normal and the tangential components of stress acting on a fault plane with dip $\beta$ and strike $\phi^\prime$. Differently from the classical definition of strike $\phi$ (angle between the intersection of the fault plane with the horizontal plane and the north direction), in the present analysis the strike is intended as the angle between the intersection of the fault plane with the horizontal plane, $\phi^\prime$, and the principal direction of stress $x$}
    \label{fig:stress_scheme}
\end{figure}

The stress on a fault plane varies with lithostatic pressure and tectonic stress. To estimate the normal and shear stresses shown in \autoref{fig:stresses}, we have assumed that $\sigma_{xx}$, $\sigma_{yy}$ and $\sigma_{zz}$ are the principal stresses (\autoref{fig:stress_scheme}), with the vertical stress $\sigma_{yy}$ that coincides with the lithostatic pressure and the horizontal stresses $\sigma_{xx}$ and $\sigma_{zz}$ equivalent to the sum of the lithostatic pressure and the tectonic stresses. The normal and shear stresses are then calculated as 
\begin{equation}
\begin{aligned}
    \sigma &= \sigma_{ij} n_j n_i \\
    \tau &= \sigma_{ij} n_j t_i
\end{aligned}
\label{eq:cauchy}
\end{equation}
where $n_i$ and $t_i$ are the unit vectors normal and tangential to the fault plane, respectively.

The Data shown in \autoref{fig:triaxial} and \autoref{fig:stresses} indicate that the relevant pressure range of interest for this study is in the 100-800 Mpa (1-8\,kbar) interval: the effect of these pressures on the piezoelectricity of quartz crystals is discussed in the next section.

\section{Muon propagation in crystalline rocks with piezoelectricity: a theoretical model}  \label{Sec:Theory}
Quartz is a fundamental mineral found in a wide range of rocks across various geological environments, including sedimentary rocks (e.g., sandstones), metamorphic rocks (e.g., quartzites), and igneous rocks (e.g., granites).
The distribution and morphology of quartz grains within these rocks significantly influence their overall physical properties, particularly their piezoelectric behavior. 
The variable grain size, orientation, and fabric of quartz crystals within the rock matrix influence its mechanical behavior, wave propagation properties, and, in specific contexts, its response to tectonic stress. 

Whenever an external pressure P acts on a quartz crystal, the piezoelectric effect will generate an electric field $E_p$ directed along the crystal axis of amplitude $E_p$\,[MV/m] $= 50\,\cdot$\,P\,[GPa] = $5\,\cdot$\,P\,[kbar]. 

This law holds for temperatures less than $200\,^\circ$C, above which the scaling decreases and the piezoelectric effect weakens, ultimately disappearing above the Curie temperature of 576\,$^\circ$C. Assuming an average continental geothermal gradient of 25–30\,$^\circ$C/km and a surface temperature of approximately 15\,$^\circ$C, the 200\,$^\circ$C isotherm is typically reached at depths between 6.2 and 7.4 kilometers. Therefore, for maximum depths of the order of less than 10\,km – and excluding crustal domains affected by anomalously high geothermal gradients – the piezo-electric scaling law reported above should represent a reliable approximation.

Muons penetrating ordinary matter are subject to energy loss due to the combined ionization and bremsstrahlung effects, plus a diffusion of transverse momenta due to MCS. The energy loss effect is described by the Bethe-Bloch-like formula \cite{dEdX}:
\begin{equation}
\label{eq:enloss}
    -\frac{\left\langle dE\right\rangle}{dz}  = a(E) + b(E)E,
\end{equation}
where $a(E)$ represents the electronic stopping power dominated by ionization losses, and $b(E)$ accounts for radiative processes including bremsstrahlung, pair production, and photonuclear interactions. 
For high-energy muons ($E \gtrsim 100$\,GeV), the radiative term $b(E)E$ becomes comparable to $a(E)$ (being equal at the muon critical energy $E_{\mu c}$), requiring a careful evaluation of both contributions in simulations and experimental analyzes.
In SiO$_2$ the muon critical energy is $E_{\mu c} = 708.2$\,GeV. For the specific case of SiO$_2$, the dependence of \autoref{eq:enloss} versus the muon energy E is plotted in \autoref{fig:dE}. In the following, we will consider relativistic muons, i.e. with energy higher than a few GeV. Therefore, in all formulas reported below, we assume that the Lorentz factor $\gamma$ of the muons is always much larger than 1, implying that they travel with velocity v close to the speed of light in vacuum c (i.e. $\beta=v/c\simeq1$). As a consequence of this assumption, the muon average life-time (which is 2.2\,$\upmu$s in its own reference frame) is largely enhanced by the relativistic time dilation: the average length traveled by muons at 1\,TeV of energy is about 6000\,km, while at 1\,GeV it reduces down to 6\,km, a length considered compatible with the requirement for muons to traverse a rock thickness of the order of one kilometer underground. 

Furthermore, for muon energy above 1\,GeV, the difference in the behavior of negative and positive muons in propagation through solid matter is negligible.

Transverse momentum diffusion arises primarily from MCS off atomic nuclei and electrons. 
This stochastic process leads to angular broadening of the muon trajectories described by the Molière angle formalism \cite{Moliere47}; the angular deviation of ultra-relativistic charged particles traversing a material follows a Gaussian distribution with a root-mean-square (RMS) angle given by:
\begin{equation}
\label{eq:moliere}
\theta_0 = \frac{13.6}{c p} \sqrt{\frac{L}{L_0}} \left[1 + 0.038 \ln\left(\frac{L}{L_0}\right)\right],
\end{equation}
where $\theta_0$ is the Molière angle, $p$ is the particle momentum, $L$ is the material thickness, and $L_0$ is the radiation length of the material \cite{Borozdin2003}. 
The term $\ln(L/L_0)$ accounts for higher-order corrections in the scattering process.  

The Molière angle $\theta_0$ depends critically on the atomic number $Z$ of the material and its density, as $L_0$ decreases rapidly with increasing $Z$.  

While the longitudinal energy loss is deterministic and well described by \autoref{eq:enloss}, the transverse diffusion introduces probabilistic fluctuations in the muon trajectory, which must be modeled statistically.

The interplay between these two phenomena determines the muon range and lateral spread in matter.
In crystalline rocks, differently from other kinds of materials, muons will be subject also to the piezo-electric field generated by the pressure acting on the embedded quartz crystals.
We will consider here only the so-called statistical piezoelectricity in rocks, as discussed in \cite{bishop1981piezoelectric}. 

Statistical piezo-electricity is due to a random configuration of quartz crystals in rocks and results in a purely random walk process to any charged particle traversing the rock layer. 
J.R. Bishop showed by experimental measurements that there are also special kinds of rocks, specifically mylonites, presenting a coherent configuration of piezoelectricity (see Figure 4 in \cite{bishop1981piezoelectric}). The particular organization of the piezo-electric field in mylonites greatly enhances the deflection of a charged particle crossing them. The quantitative evaluation of this coherent piezoelectric enhancement needs future experimental measurements, as those shown in the proof-of-principle experiments of second phase, performed with different kinds of multi-crystalline rock samples and discussed in section VI (sub-section B).  

Crystals can indeed be represented as simple diodes with an electrostatic field given in module by $E_p$ (see formula above) and whose direction is randomly distributed over the whole solid angle, following the casual axis of the quartz crystals distributed inside the granite (or granite-like) rock matrix. The effect of statistical piezoelectricity on the muon propagation is modeled by imparting to the particles crossing the quartz crystals small perturbative kicks due to the piezoelectric field excited by the active pressure on the rock. 

As an example of this simplified scheme, a 1\,cm wide quartz crystal subject to a pressure of 1\,kbar will present an applied voltage of 50\,kV corresponding to an electric field of 50\,kV/cm = 5\,MV/m  and directed along its symmetry axis.
If the crystal axis is aligned with the x or the y-axis  (the muon propagates along z), after traversing the crystal, the muon transverse momentum $p_r=\sqrt{p_x^2+p_y^2}$ will be suddenly changed in modulus by an amount equal to $|\Delta p_r| = e E_p L_c/c$ , where $L_c$ is the thickness of the quartz crystal.
For a 100\,GeV muon crossing a 1\,cm quartz crystal under 1\,kbar of pressure, the angular kick imparted by the piezoelectric field will be about 50\,$\upmu$rad.
In order to describe the propagation of the high-energy muons, we adopt a step-wise description of the motion using a thin-lens approximation: in the impact with the quartz crystal, the muons change only the momentum, while, instead, the transverse position (x, y) changes in the drifts  inside the rock between two consecutive quartz crystals  by an amount proportional to the accumulated transverse angle ($r^\prime$) .
Because of the randomly distributed direction of the piezoelectric field active inside the quartz crystals, we will treat the momentum kicks via a random walk approach/description: the longitudinal momentum $p_z$, being dominated by its very initial large value (as typical for multi-hundreds GeV muons), will stay on average unaffected by the random walk process, and decreases only due to ionization and radiative losses, while the transverse momentum $p_r$ of the muon will follow a 2D random walk behavior.

According to the well-known 2D random walk formula, we can write the average accumulated transverse momentum as $p_r(z)= E_p\sqrt{z_cL_c}$ ($p_r$ expressed in MeV/$c$ and $E_p$ in MV/m), assuming the initial condition $p_r=0$ at $z=0$ at the beginning of the propagation through the crystalline rock layer. Here, $z_c=zf_c$ , where $f_c$ represents the filling factor of quartz crystals inside the hosting crystalline rock, that is, how much rock thickness crossed by the muons is filled by quartz ($0< f_c <1$). $z_c$ is therefore the active length where the transverse momentum actually undergoes the 2D random walk process.  In order to derive an approximate analytical formula, the average muon trajectory angle $\sigma_{r^\prime}=p_r(z)/p_z(z)$ can be calculated by assuming a linear decrease of the muon longitudinal momentum in the propagation through the rock,  corresponding to the first order approximation of the solution of Eq. 3,   $p=p_0-\delta z$, where $\delta$ is the longitudinal momentum loss evaluated in $p_0$.

Integrating the trajectory angle 
\begin{equation}
\label{eq:sigmaprime}
\sigma_{r^\prime}(z)=\frac{2}{\pi}\frac{E_p\sqrt{zL_c f_c}}{p_0-\delta z} 
\end{equation}

over the z coordinate, we can derive the average muon displacement $\sigma_r(z)$ as:
\begin{equation}
\label{sigmar}
\sigma_r(z)=\frac{4}{\pi}\frac{E_p\sqrt{L_c}}{\delta}\left(\sqrt{\frac{p_0}{\delta}}\operatorname{tanh}^{-1}\sqrt{\frac{\delta z}{p_0}}-\sqrt{z}\right).    
\end{equation}

As an example referred to the data of \autoref{fig:depth_magnitude_distributions}, taking a 1\,TeV muon beam traversing a 1\,km thick SiO$_2$ rock layer, with an average of 0.9\,GeV energy loss per m, filled with 1\,cm-thick quartz crystals ($L_c=1$\,cm, $f_c=1/2$) subject to a 10\,kbar pressure generating 50\,MV/m piezo-electric field, the average trajectory angle at the end of the rock layer (exit energy 100\,GeV) will be $\sigma_{r^\prime}=0.7$\,mrad for an average muon displacement of  $\sigma_r=21$\,cm.
A second relevant example, which will be further analyzed in \autoref{sec:MuAEGIS} by detailed numerical simulations, considers 500\,GeV muons crossing 600\,m of rock with an average energy loss of $0.75$\,GeV per meter, and a 4\,kbar of tectonic applied pressure generating a piezoelectric field of 20\,MV/m in quartz crystals assumed to be 2\,cm thick ($L_c=2$\,cm), with a filling factor $f_c=2/3$. The muon beam exits, in this case,  the 600\,m long rock layer at 50\,GeV energy with $\sigma_{r^\prime}=0.72$\,mrad and  $\sigma_r=11$\,cm.

Since $\sigma_{r^\prime}$ and $\sigma_r$ are respectively angular and position rms values in the muon phase space distribution generated by the piezoelectric random walk (PRW, from now on) diffusion process, we can compute the transverse emittance generated by the diffusion process as:
\begin{equation}
\label{emittance1}
\begin{split}
\varepsilon_\text{\tiny PRW}(z) &= \sigma_{r^\prime}(z)\sigma_r(z)/(2\sqrt{5}) = \\ &= \frac{4E_p^2L_c\sqrt{z}}{\pi^2\sqrt{5}\delta\left(p_0-\delta z\right)}\left(\sqrt{\frac{p_0}{\delta}}\operatorname{tanh}^{-1}\sqrt{\frac{\delta z}{p_0}}-\sqrt{z}\right).
\end{split}
\end{equation}
The normalized emittance will then be:  
\begin{equation}
\label{emittance2}
\begin{split}
\varepsilon_{n,\text{\tiny PRW}}(z) &= \left(p_0-\delta z\right)\varepsilon_\text{\tiny PRW}(z)/mc .
\end{split}
\end{equation}
where m=105.7\,MeV is the muon mass. In the first example mentioned above (1\,TeV muons), the emittance dilution suffered by the beam at the end of the 1\,km rock crossing will be huge, 0.031\,m$\cdot$rad, while in the second example (500\,GeV muons) will be 0.008\,m$\cdot$rad after 600\,m.

Following a similar approach, we calculate the Molière angle (\autoref{eq:moliere}) for the case of SiO$_2$ with relativistic muons, neglecting the high-order correction term $\ln(L/L_0)$. The rms angle of the muon beam is given by (the suffix MCS stays for Multiple Coulomb Scattering):

\begin{equation}
\label{Moliere1}
\theta_{\text{\tiny MCS}} = \frac{41.9 \sqrt{z}}{\left(p_0 - \delta z\right) [ \frac{\text{MeV}}{\text{c}}]},
\end{equation}
that, integrated over the z coordinate, gives an approximate expression of the muon beam rms spot size $\sigma_\text{\tiny MCS}$

\begin{equation}
\label{spot}
\sigma_{\text{\tiny MCS}} = \frac{83.8}{\delta} \left( \sqrt{\frac{p_0}{\delta}} \tanh^{-1} \sqrt{\frac{\delta z}{p_0}} - \sqrt{z} \right).
\end{equation}
The normalized increase in emittance due to the MCS mechanism is given by:

\begin{equation}
\label{emittance3}
\varepsilon_{n,\text{\tiny MCS}} = \theta_{\text{\tiny MCS}}\sigma_{\text{\tiny MCS}}\left(p_0-\delta z\right)/(2\sqrt{5}mc).
\end{equation}
Combining the two effects, MCS and PRW, we can express the rms spot size of the muon beam at the exit (assuming $\Delta \ll 1$) as $\sigma_\text{\tiny TOT}$,

\begin{equation}
\label{eq:sigmatot}
\sigma_{\text{\tiny TOT}} = \sigma_{\text{\tiny MCS}} \left(1 + \frac{1}{2} \frac{\sigma_r^2}{\sigma_{\text{\tiny MCS}}^2}\right)=\sigma_{\text{\tiny MCS}} (1 + \Delta),
\end{equation}
where $\Delta$ is given by:
\begin{equation}
\label{eq:Delta}
\Delta = \frac{\sigma_{\text{\tiny TOT}}}{\sigma_{\text{\tiny MCS}}} - 1 = 1.15 \cdot 10^{-4} \, E_p^2 \, L_c
\end{equation}
and represents the relative enhancement of the beam spot size due to piezoelectric effects driven by the applied tectonic pressure.
This parameter is the real signature of the tectonic time evolution toward a seismic event: monitoring $\Delta$ over time will allow us to follow in real time how a fault is evolving its status towards a rupture event.

By using the previous example of 500\,GeV muons crossing a 600\,m thick layer of SiO$_2$ with $E_p=20$\,MV/m, $L_c=2$\,cm and $\delta=0.75$\,GeV/m, we find: $\theta_{\text{MCS}}=25$ mrad, $\sigma_{ \text{TOT}}=2.5$\, m, $\epsilon_{n, \text{MCS}} =5.4$\,m$\cdot$rad.
Since the contribution of piezoelectric random-walk effect is $\sigma_r=11$\,cm (\autoref{eq:sigmatot}), we find that $\Delta= 0.001$, defined as in \autoref{eq:Delta}, is  0.1\,\% of the MCS effect, corresponding, in absolute quantities, to an increase of about 2.5\,mm of the unperturbed beam spot size of 2.5\,m.

MCS is the dominant effect, while PRW adds a perturbative effect both in angle and beam spot size.
However, it is worth noting that the final values of the muon beam emittance after traversing km-long rock layers are quite huge, i.e. in the range of meter.radians,  several orders of magnitude larger than the typical small emittances requested by muon collider scenarios (in the range of tens of mm$\cdot$mrad).
As discussed further in the next section, this fact implies that the accelerator foreseen to provide the muon beam requested for the ERMES technique is not supposed to provide small emittance muon beams, greatly releasing the challenges inherent in the generation, transport and manipulation of the muon beam from the source up to the injection into the rock layer across the fault area to be monitored. 
The huge phase-space dilution applied to the muon beam by the stochastic diffusive process of the combined MCS and PRW effects makes the exit muon beam parameters (rms angle and spot size) almost independent of the initial beam optics at the entrance into the rock layer.

Using the last two equations, we can evaluate the total number of muons per pulse requested for a good imaging of the PRW effect: the parameter $\Delta$ represents indeed the relative thickness of the halo generated by PRW over the transverse beam distribution, which is dominated by the MCS effect.
If $\rho_0$ is the beam density averaged over the entire muon beam spot at the exit of the rock layer and $\sigma_\text{TOT}$ the spot rms dimension, the ratio between the number of muons comprised in the halo $N_\Delta$ and the total number of muons in the beam $N_\text{TOT}$ is given by $N_\Delta/N_\text{TOT}=2\Delta$.
The total number of muons as a function of the muons in the halo, taking into account the given expression for $\Delta$, is  $N_\text{TOT}=4300\cdot N_\Delta/(E_p^2L_c)$.
If we consider the most unfavorable case, for which the requested number of muons in the halo is $N_\Delta=10^3$,  so to achieve a good resolution, and assume $E_p$=4\,MV/m, corresponding to a weak tectonic pressure of 0.8\,kbar, with 1\,cm thick quartz crystals ($L_c=0.01$), we obtain $N_\text{TOT} = 2.7 \times 10^7$.
This number corresponds to a quite modest bunch population for any accelerator system expected to generate and accelerate the muon beam up to final energy of 500\,GeV.

\section{Numerical simulations of MCS and PRW with a new framework: MuAEGIS}
\label{sec:MuAEGIS}
The MuAEGIS (Muon Underground Active Earthquake Genesis Investigation Software) code models relativistic charged particle transport through piezoelectric crystalline structures under extreme tectonic stress conditions. 
Its primary scientific objective is to quantify how pressure-induced electric fields modify beam dynamics compared to baseline MCS and energy loss processes. 

\begin{figure}[htbp]
    \centering
    \includegraphics[width=1\linewidth]{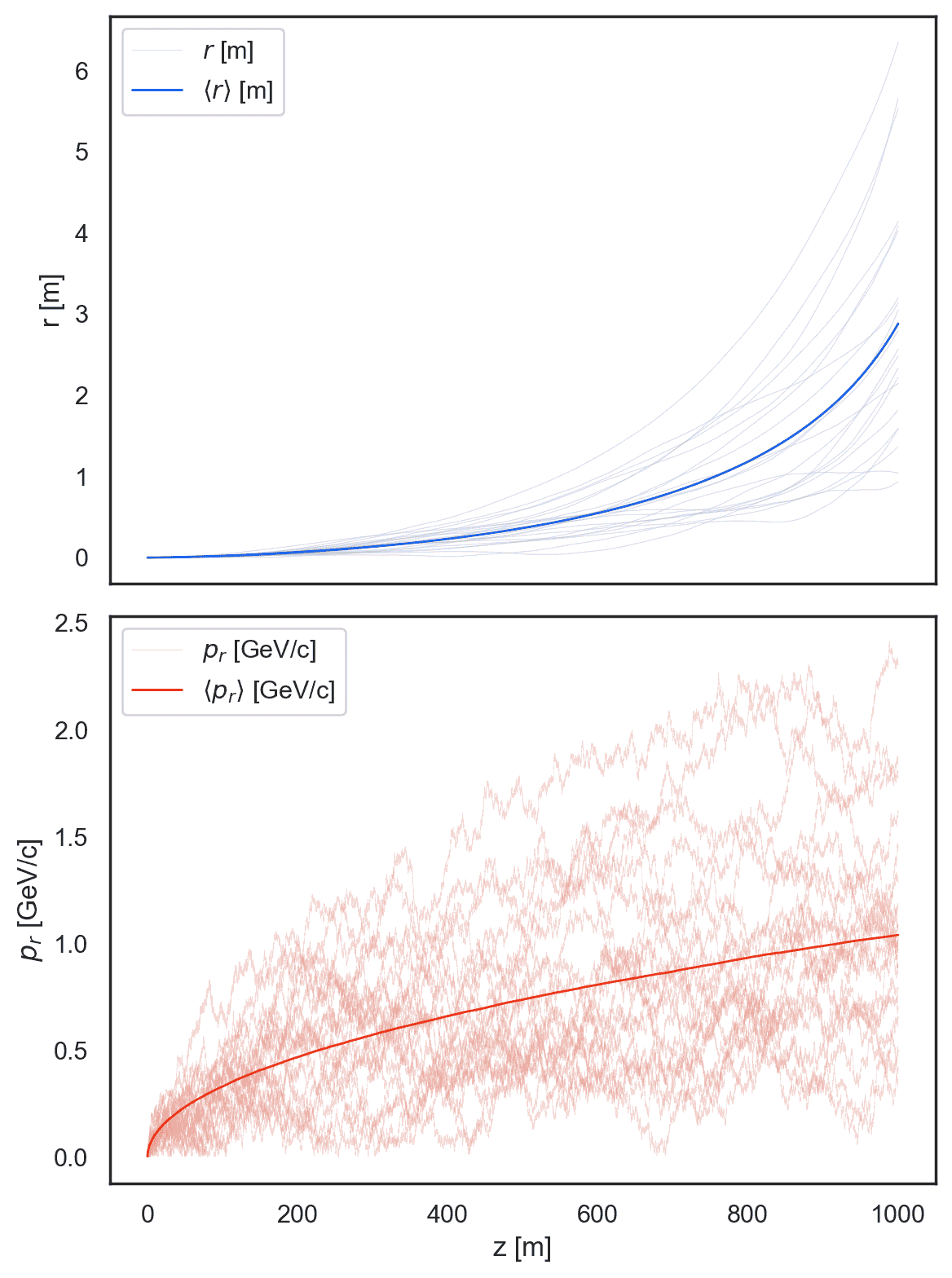}
    \caption{Tracking of $2\times10^4$ muons along 1\,km of crystalline rock.
    The path is composed by alternated layers of quartz and simple rock each 1\,cm long.
    Radial coordinate $r$ in light blue and radial momentum $p_r$ in light red are displayed for a representative subset of the bunch (20 particles). 
    The average value of these quantities are shown respectively in blue and red.}
    \label{fig:tracking}
\end{figure}

The simulation combines three fundamental interactions: energy loss through ionization and radiation losses, MCS, and PRW – with particular emphasis on their collective impact on transverse beam dimensions.

Particle trajectories evolve through sequential momentum updates governed by:
\begin{itemize}
    \item \textbf{Energy loss}: calculated using cubic interpolation of data from comprehensive tables on muon stopping power and range, in the cases shown in this work, we consider SiO$_2$ rocks, providing $dE/dz$ values across the 1\,MeV–1\,TeV energy range \cite{Groom2001Muon, ParticleDataGroup:2024cfk}.

    \item \textbf{Multiple Coulomb Scattering}: Evaluation of the Molière scattering angle following \autoref{eq:moliere} applying it with a Monte Carlo approach in a 3-dimensional reference system. 
    The code works under the paraxial approximation ($p_z \gg p_r$) to ensure faster computation as the scattering angles are typically below the microradians in the cases considered.
    
    \item \textbf{Piezoelectric Random-Walk (PRW) steering}: momentum kicks by quartz crystals of length $L_c$ given by: $\Delta p = eE L_c/c$.
    The field E is subject to field suppression when $t>200\ \,^\circ$C. Its direction is stochastically distributed over the entire solid angle.
\end{itemize}

The code initializes particles using truncated Gaussian distributions in the transverse phase space or imports externally generated beam distributions to emulate realistic profiles from tracking simulation tools.
In pressure-off simulations ($P = 0$\,kbar), the piezoelectric terms vanish, and only stochastic scattering processes remain active.

The primary comparison metric—beam envelope evolution, is obtained through statistical analysis of the phase space of the tracked particles as they traverse $N$ quartz crystals, alternated with rock layers of adjustable length. The intermediate rock layers are assumed to be neutral from the point of view of piezoeletricity, with the same density and radiation length as SiO$_2$.
The simulation evaluates, layer by layer, both stochastic MCS and piezoelectric steering effects via continuous updates of the $\beta\gamma$ factor, and considers three distinct propagation scenarios: MCS only, PRW only, and the combined action of MCS and PRW.

\begin{figure}[htbp]
    \centering
    \includegraphics[width=1\linewidth]{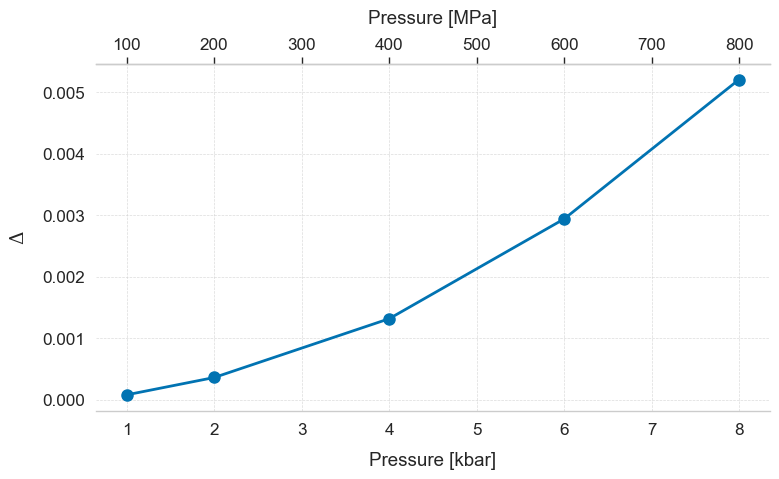}
    \caption{Relative enhancement $\Delta$ of the beam envelope as a function of the applied pressure in quartz crystals, ranging from 1 to 8\,kbar (100 to 800\,MPa).  
    The simulation is based on an initial muon bunch with normalized emittance $\varepsilon_n = 50$\,mm$\cdot$mrad.  
    A clear nonlinear increase of $\Delta$ is observed, with values growing from $\sim10^{-4}$ at 1\,kbar to over $5\times10^{-3}$ at 8\,kbar.  
    This trend reflects the strengthening of the piezoelectric-induced transverse momentum diffusion as the stress in the crystal increases.  
    The upper x-axis reports the equivalent pressure in MPa for reference.}
    \label{fig:Deltas}
\end{figure}
MuAEGIS is designed such that each particle propagates through a unique configuration of quartz crystal orientations, reproducing the stochastic nature of particle trajectories without requiring storage of the full 3D geometry of the environment, which would require excessive memory resources.

As a result, the radial momentum of each particle undergoes stochastic evolution along its trajectory.  
However, the average behavior of the bunch's radial momentum follows the well-known scaling law of two-dimensional random walks: $\sqrt{\langle p_r^2 \rangle} \propto \sqrt{N}$, where $N$ is the number of crystals traversed.  
In \autoref{fig:tracking}, we present the statistical behavior of an ensemble of $2\times 10^4$ muons. 
The two panels show, respectively, the average radial position of the particles tracked over 1\,km of crystalline rock (top panel, in blue) and the average radial momentum (bottom panel, in red).  
To illustrate the stochastic nature of individual trajectories as well as the emergence of order through statistical averaging, a representative subset of 20 muons is displayed in light blue (top) and light red (bottom).

\begin{figure}[htbp]
    \centering
    \includegraphics[width=1\linewidth]{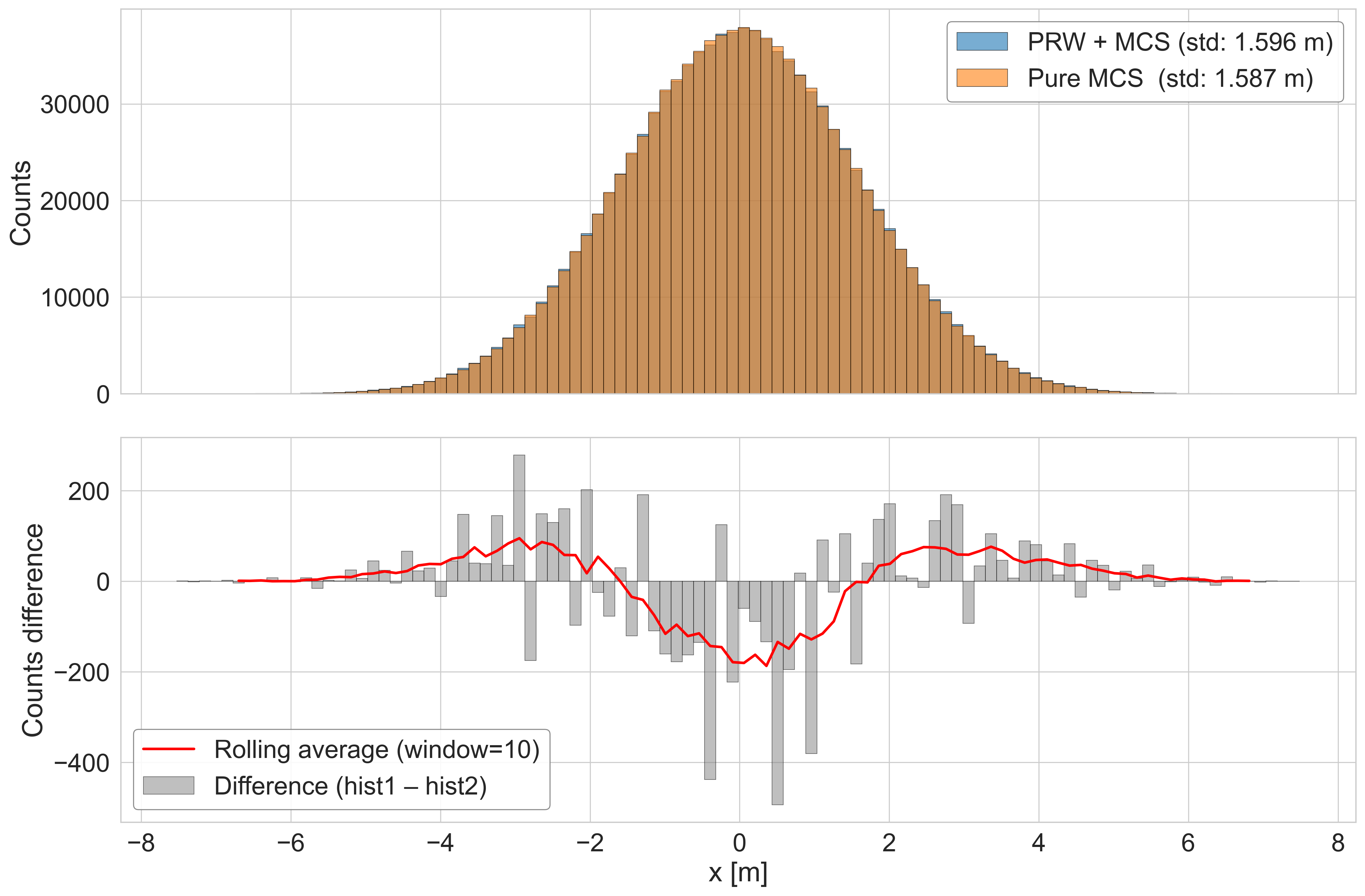}
    \caption{Comparison of muon beam profiles with and without PRW effect at 8\,kbar pressure. 
    Upper plot: Overlaid histograms of the muon arrival positions ($x$) on the detector plane for the PRW\,+\,MCS case (blue) and the pure MCS case (orange). 
    The two distributions appear nearly identical in terms of standard deviation ($\sigma \approx 1.59$\,m). 
    Lower plot: Bin-by-bin difference between the two histograms, highlighting a net redistribution of counts from the central region to the tails in the PRW case. 
    A rolling average (red line, window=10) emphasizes the trend.}
    \label{fig:Hist Comparison}
\end{figure}
MuAEGIS considers also temperature-dependent effects, parametrized through a $t_{\text{coeff}}$ suppression factor and simulates the pressure-dependent piezoelectric field calculating it from the average pressure value, enabling rapid evaluation of geological parameter variations.
Numerical implementation combines NumPy-based matrix operations \cite{harris2020array} with multiprocessing across 64 CPU cores for efficient phase-space sampling. 
Statistical convergence is verified through ensemble averaging and making use of a high number of particles, $N>10^6$ typically, due to the parallel nature of the code. 

In order to test the code predictions, we simulated the case of $10^6$ muons arriving on a detector after a propagation through 600\,m in granite rock.
The quartz crystals contained in the granite aggregate have an average length of 2\,cm and are separated by 1\,cm of neutral rock from the point of view of piezoelectricity.  
The muons propagate in the rocks with an initial kinetic energy of 500\,GeV.
The rock is parametrized considering SiO$_2$ with density $\rho = 2.2$\,g/cm$^3$ a radiation length $\text{L}_0^{\text{SiO}_2} = 12.3$\,cm.
We consider a total pressure (considering lithostatic and tectonic pressure) $ P=4$\,kbar = 400\,MPa, representing nearly the maximum achievable value near a fault zone prior to the occurrence of an earthquake event, and a temperature of $T < 200\,^\circ$C, as typical of shallow fault systems.

We performed several simulations of this system under different initial conditions.
Initially, we simulated particles with identical starting parameters; subsequently, we propagated particle distributions representing a more realistic phase-space configuration.
In all simulations, each particle traveled through a different configuration of quartz crystal orientations, mimicking the stochastic nature of the paths that muons are expected to follow.

The goal of the preliminary study using identical initial conditions ($x = y = z = 0$, $p_x = p_y = 0$, and $p_z = 500$\,GeV/$c$) was to estimate the expected PRW effect on the beam envelope and divergence. 
As a result, we obtained, on the detector plane located 600\,m downstream from the starting point, a muon bunch with RMS transverse dimensions $\sigma_x \approx \sigma_y$ and RMS transverse momenta $\sigma_{p_x} \approx \sigma_{p_y}$.
As expected, the beam centroids remained close to zero, and the system preserved cylindrical symmetry. For this reason, we adopt the radial coordinates in the transverse plane, referring to position and momentum as $r$ and $p_r$, respectively.

\begin{figure*}
    \centering
    \includegraphics[width=1\linewidth]{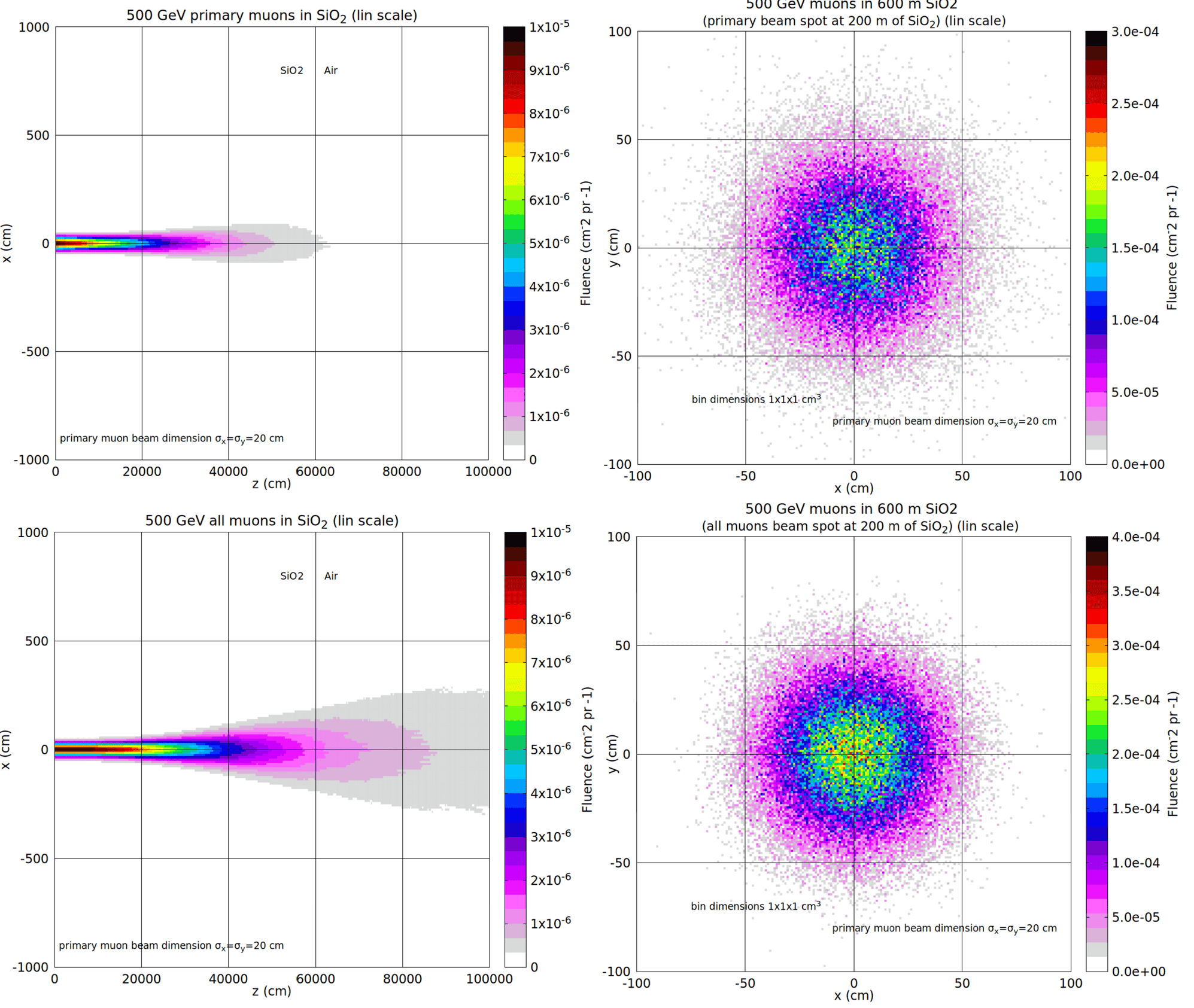}
    \caption{FLUKA results showing muon beam fluence (on the left) and transverse density distributions (on the right). 
    Upper diagram on the left side shows the fluence of only primary muons traversing 600\,m of SiO$_2$ and 400\,m of air. 
    Lower diagram on the left side shows the fluence of all muons (primary, secondary, next generations). 
    Upper diagram on the right shows the muon beam transverse density distribution of only primary muons after 200\,m of rock, while the lower diagram shows all muons.}
    \label{fig:FluenceSpots}
\end{figure*}
We also performed simulations of the same system while selectively disabling either the MCS effect (to isolate the pure PRW contribution) or the PRW effect (to isolate the pure MCS contribution).
The resulting transverse beam parameters from these simulations are summarized in \autoref{tab:SinglePart}.
\begin{table}[htbp]
\centering
\begin{tabular}{l c c c}
\toprule
 & $\sigma_r$  & $\sigma_{p_r}$& $\varepsilon_{nr}$\\
 & [m] & [MeV/$c$] & [m$\cdot$rad]\\
\midrule
Pure PRW         & 0.114  & 46.183   & 0.013 \\
Pure MCS         & 2.228  & 899.259  & 4.784 \\
PRW\,+\,MCS        & 2.231  & 900.497  & 4.797 \\
\bottomrule
\end{tabular}
\caption{Transverse beam parameters obtained from single-particle tracking simulations at 600\,m from the starting point. The table shows the RMS transverse radius $\sigma_r$, the RMS transverse momentum $\sigma_{p_r}$, and the normalized transverse emittance $\epsilon_{nr}$ for three scenarios: with only piezoelectric random walk (PRW), with only multiple Coulomb scattering (MCS), and with both effects combined.}

\label{tab:SinglePart}
\end{table}

\begin{figure*}
    \centering
    \includegraphics[width=1\linewidth]{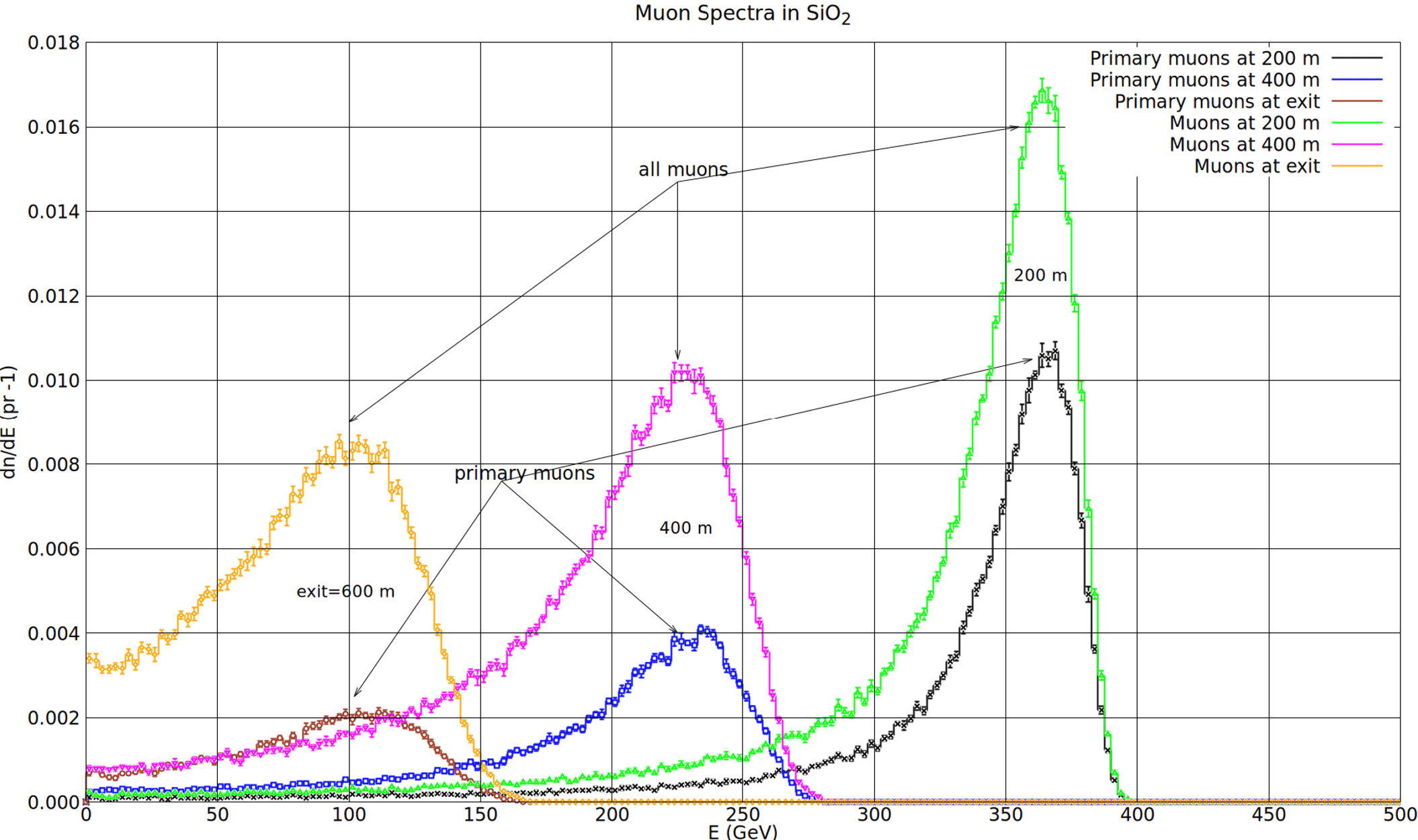}
    \caption{Energy spectra of primary and all muons at various locations through the rock traversal, i.e. after 200\,m, 400\,m, and at the rock layer exit at 600\,m.}
    \label{fig:FlukaSpectra}
\end{figure*}
The average beam longitudinal momentum at the detection plane is $\langle p_z \rangle \approx 49.9$\,GeV/$c$, leading to an RMS beam divergence of $\sigma_{r'} \approx 12.8$ mrad.

Subsequently, we simulated more realistic bunch distributions, assuming a two-dimensional Gaussian profile with $\sigma_x = \sigma_y = 20$\,cm, truncated at $5\sigma$, and an initial normalized emittance of $\varepsilon_{n,x} = \varepsilon_{n,y} = 50$\,mm$\cdot$mrad.  
The results of the simulations for the three configurations—pure PRW, pure MCS, and combined PRW+MCS are reported in \autoref{tab:50mmmrad}.

\begin{table}[htbp]
\centering
\begin{tabular}{l c c c}
\toprule
 & $\sigma_r$  & $\sigma_{p_r}$& $\varepsilon_{nr}$\\
 & [m] & [MeV/$c$] & [m$\cdot$rad]\\
\midrule
Pure PRW         & 0.305  & 46.187   & 0.088 \\
Pure MCS         & 2.243  & 898.059  & 5.068 \\
PRW\,+\,MCS        & 2.246  & 899.271  & 5.081 \\
\bottomrule
\end{tabular}
\caption{Transverse beam parameters from simulations using a realistic initial phase-space distribution, with a round RMS spot size of 20\,cm and an initial normalized transverse emittance of 50\,mm$\cdot$rad. Results are shown for three configurations: only piezoelectric random walk (PRW), only multiple Coulomb scattering (MCS), and both effects combined.}
\label{tab:50mmmrad}
\end{table}
The same procedure was applied to a beam with identical transverse size but degraded quality, characterized by a higher initial normalized emittance ($\varepsilon_{n,x} = \varepsilon_{n,y} = 500$\,mm$\cdot$mrad). The corresponding results are shown in \autoref{tab:500mmmrad}.

\begin{table}[htbp]
\centering
\begin{tabular}{l c c c}
\toprule
 & $\sigma_r$  & $\sigma_{p_r}$& $\varepsilon_{nr}$\\
 & [m] & [MeV/$c$] & [m$\cdot$rad]\\
\midrule
Pure PRW         & 0.305  & 46.223   & 0.088 \\
Pure MCS         & 2.243  & 898.451  & 5.072 \\
PRW\,+\,MCS        & 2.246  & 899.591  & 5.084 \\
\bottomrule
\end{tabular}
\caption{Transverse beam parameters from simulations using a realistic initial phase-space distribution with a round RMS spot size of 20\,cm and an initial normalized transverse emittance of 500\,mm$\cdot$rad. The table reports the RMS transverse radius $\sigma_r$, RMS transverse momentum $\sigma_{p_r}$, and normalized transverse emittance $\epsilon_{nr}$ for the cases including only piezoelectric random walk (PRW), only multiple Coulomb scattering (MCS), and the combined effect.}
\label{tab:500mmmrad}
\end{table}
As the results indicate, increasing the emittance by a factor of ten does not significantly affect the outcome, since the beam quality degradation due to PRW and MCS largely outweighs the initial beam parameters.  
This demonstrates that low emittance muon beams similar to those requested by muon colliders are not strictly required to observe the investigated effect. 
A muon beam normalized rms transverse emittance up to 500-1000\,mm$\cdot$mrad looks at all adequate for this technique. 
It should be also noted that the muon beam envelope evolution along the km-scale propagation length, for the large typical spot size considered at injection (20 cm rms), even with very large normalized emittances of 1000 mm.mrad, would be almost parallel in vacuum, since the associated optical beta-function would be in the range of hundreds of kilometers: this implies that the calculated strong muon beam defocussing, leading the beam spot size up to a couple meters in size at the exit from the 600 m rock layer (starting from 20 cm at injection), is only determined by the scattering processes, not by the linear optical beam envelope behavior.

Conversely, variations in the initial beam size do affect the emittance growth, as is well known also in the Molière angle formalism.  
This behavior is expected, as the phase space diffusion induced by the random walk processes acts primarily on the transverse momentum, enlarging the beam phase space area (and hence the emittance) in proportion to the initial spatial extent of the bunch.

The results obtained with MuAEGIS for realistic muon bunches show good agreement with the theoretical predictions presented in \autoref{Sec:Theory}, with discrepancies on the order of 10-20\% on average.  
The beam envelope values extracted from all the simulated distributions correspond to a relative enhancement of about $\Delta \approx 0.0013$, consistent with theoretical expectations.

Subsequently, MuAEGIS was employed to study the scaling behavior of the $\Delta$ factor with increasing pressure, simulating conditions that may occur in the lead-up to an earthquake event.  
We simulated a muon beam with initial normalized emittance $\varepsilon_{n,x} = \varepsilon_{n,y} = 50$\,mm$\cdot$mrad under pressures of 1, 2, 4, 6, and 8\,kbar (i.e., 100, 200, 400, 600, and 800\,MPa), and analyzed the evolution of the $\Delta$ parameter.  

The data reported in \autoref{fig:Deltas} reveal a nonlinear dependence of the $\Delta$ factor on the applied pressure.  
At low pressures (1–2\,kbar), the enhancement remains modest, below $10^{-3}$, but increases rapidly for higher stress levels, reaching values of the order of $5 \times 10^{-3}$ at 8\,kbar.  
This behavior supports the hypothesis that piezoelectric interactions in the crystalline medium become increasingly effective at steering the beam as the internal stress rises, which is consistent with the expected behavior near seismic fault activation.  
The observed trend could potentially be modeled by a power law or exponential growth, suggesting a strong coupling between mechanical stress and transverse phase space diffusion.

For the case corresponding to an 8\,kbar pressure, we evaluated the expected detectable modification of the muon distribution induced by the PRW effect, compared to the baseline case in which only pure MCS is present.
To this aim, we compare the histograms of the horizontal coordinates ($x$) of particle arrivals at the detector plane for the two scenarios, overlaid in the upper plot of \autoref{fig:Hist Comparison}.
The difference is barely noticeable, as the RMS values are very close: $\sigma_{x_{\text{PRW+MCS}}} \approx 1.596$\,m and $\sigma_{x_{\text{MCS}}} \approx 1.587$\,m. 
For a radially symmetric distribution, it is useful to recall that $\sigma_r = \sqrt{2}\sigma_x$.
The distinctive behavior becomes clearer when subtracting the histograms bin-by-bin and applying a rolling average to reduce statistical fluctuations, as shown in the bottom plot of \autoref{fig:Hist Comparison}. 
The result reveals that the PRW effect leads to an enhanced redistribution of particles from the center toward the tails of the distribution, compared to the stress-free (pure MCS) case.

\begin{figure}[htbp]
    \centering
    \includegraphics[width=1\linewidth]{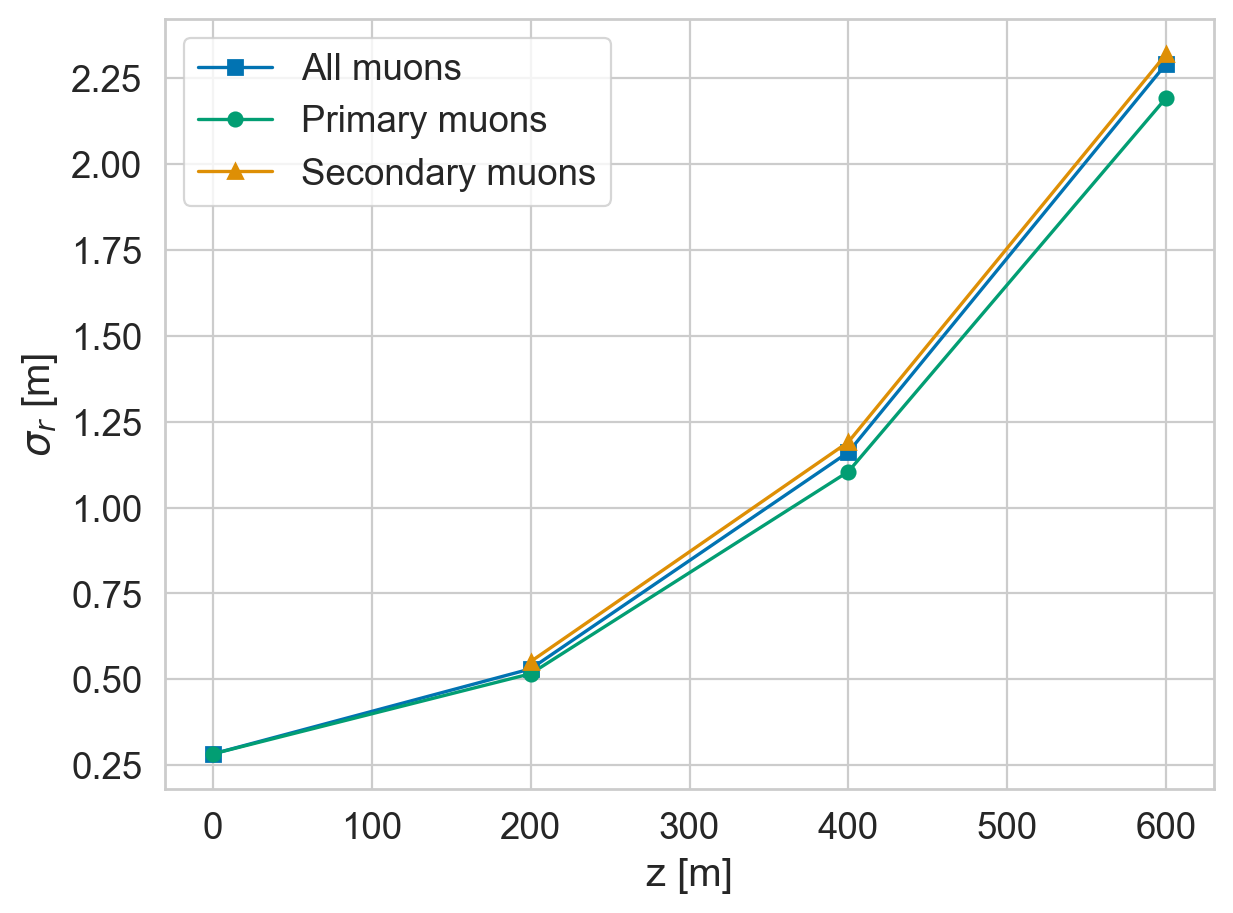}
    \caption{Rms radial beam spot size evolution of the muon beam simulated with FLUKA, considering only primary muons (green line), only secondary muons (orange line) and all muons (blue line).}
    \label{fig:Envelopes}
\end{figure}

Since MuAEGIS is based on a semi-analytical approximation, and neglects also all interactions of the primary muons with matter, which in turn produce secondary particles, we conducted extensive simulations with the Monte Carlo code FLUKA in order to assess the survival rate of muons in traversing the km-long rock layer, the overall effect on their beam envelope, and their energy degradation leading to the energy spectrum exhibited at the exit (this information is not provided by MuAEGIS due to its inherent approximation schematics). 
FLUKA is not presently equipped with the capability to describe the piezoelectric induced random walk process, therefore, its results should be compared to MuAEGIS' ones carried on with just MCS on, and PRW off. 
Using the same set of parameters illustrated in previous paragraphs, we found that the 500\,GeV primary muons survive only by a fraction of 19\% at the exit of 600\,m thick SiO$_2$ layer, but lots of secondary muons of high energy are generated by primary muon interactions with nuclei and atomic electrons, so at exit we collect secondary muons together with the surviving primary ones at an overall rate nearly equal to 80\% of the total number of primary muons launched into the rock. 
To illustrate this result, we plot in \autoref{fig:FluenceSpots} the muon fluence evolution, expressed in number of particle per cm$^2$ as predicted by FLUKA, along 600\,m of SiO$_2$ and a following 400\,m of air. 
The upper diagram on the left side shows the fluence associated to only primary muons, while the lower diagram on the left side plots the fluence of all muons, primary, secondary and all subsequent generations. A cut-off was applied to muons lower than 10\,MeV kinetic energy. 
The initial drop of primary muons is clearly visible along the first 100 meters through the rock, while the total muon fluence (top diagram) shows a much more gentle decrease, leading to the overall 80\% fraction of muons that exit the rock layer after 600\,m. The muon transverse density distributions are shown on the right side in \autoref{fig:FluenceSpots}, after 200\,m of rock thickness (only primary muons in upper diagram, all muons in lower diagram). 
The energy spectra are plotted in \autoref{fig:FlukaSpectra}, where the distribution histograms are shown after 200\,m, 400\,m, and at the rock layer exit at 600 m (only primary muon spectra and all muon spectra are indicated by the arrows). 
The muon beam rms envelope evolution ($\sigma_r$) is plotted in \autoref{fig:Envelopes}, once again separating the contributions of only primary, only secondary and all muons. 
The value of the rms radial spot size of the all muon beam is remarkably close to what evaluated by MuAEGIS (see \autoref{tab:SinglePart}, \autoref{tab:50mmmrad} and \autoref{tab:500mmmrad}).

\section{Conceptual lay-outs of high energy muon accelerators for earthquake forecast with ERMES} \label{Sec:Layout}

\begin{figure}[htbp]
    \centering
    \includegraphics[width=1\linewidth]{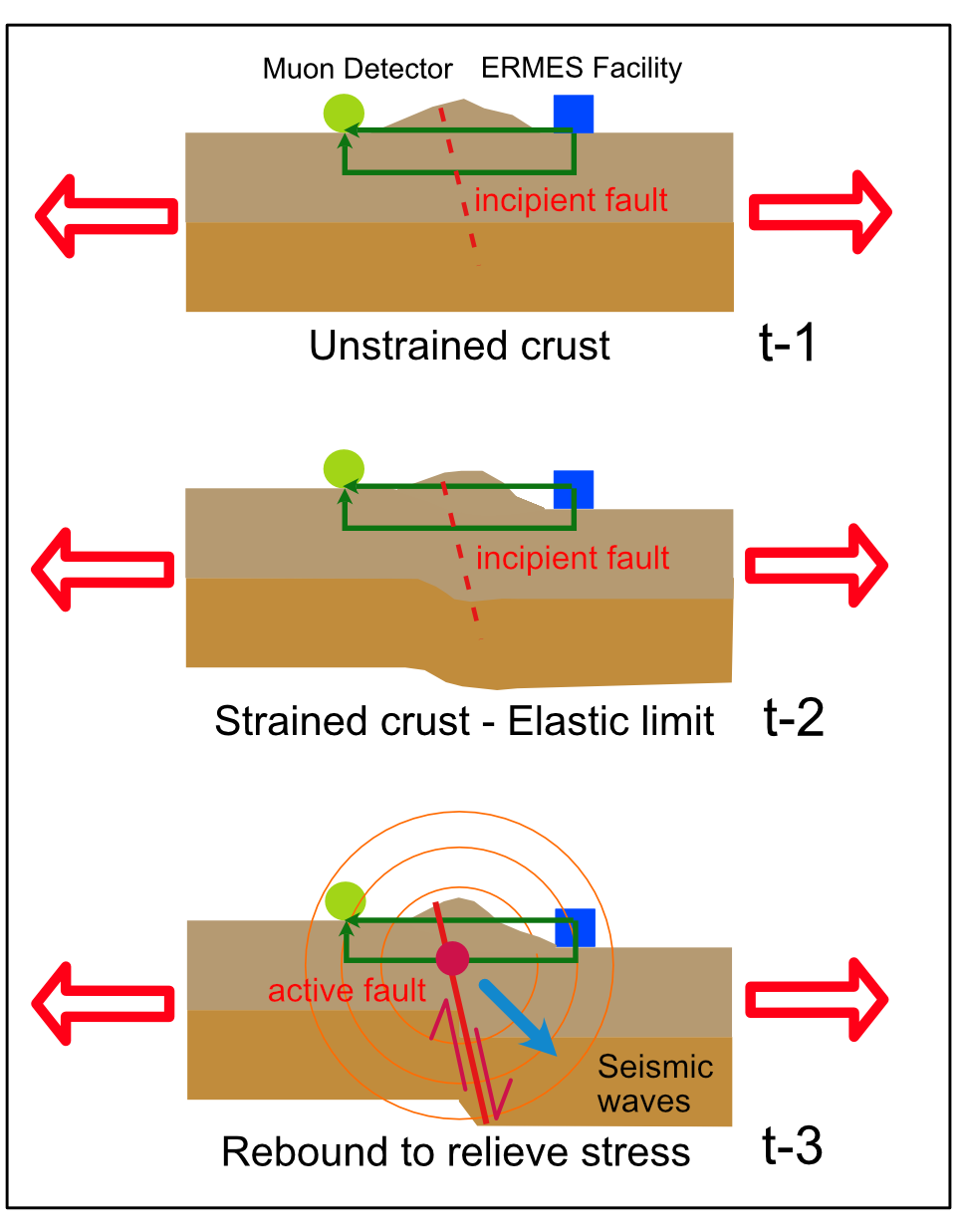}
    \caption{Conceptual layout of an ERMES facility (Earthquake Reconnaissance via Muon beam Evolution in Silicon dioxide) in a simplified geological fault scenario generated by tectonic stresses a) Unstrained crust at time 1 (t-1) with the muon beam traveling along  mostly horizontal paths between source and detector. b) Strained crust at elastic limit at time 2 (t-2) with the incipient fault. c) Active fault, producing rebound to relieve stresses (t-3). The muon beam traveling at depth is generated at ground level, directed downward, then bent 90° to traverse the rock horizontally across the fault and reach an underground detector. Red arrows represent directions of tectonic pressures}
    \label{fig:Layout}
\end{figure}

Assuming that the signal carried by the muon beam, sampling the PWR effect in traversing the km-class crystalline rock layer across a fault, can be properly measured by an adequate muon detector located right adjacent the exit from the rock, one has to consider how the conceptual schematics of the whole muon beam system may be configured. Based on the state-of-the-art angular resolution performances of silicon strip detectors, down to 20\,$\upmu$rad, as discussed in \cite{Abbiendi:2017}, we notice that the results on transverse momentum shift due to PWR, reported in \autoref{tab:SinglePart}, \autoref{tab:50mmmrad} and \autoref{tab:500mmmrad} are consistent with a request of 20 $\upmu$rad angular resolution. 
More relaxed seems the request on spatial resolution, that is in the range between a few hundreds of microns up to a few millimeters, according to Table I, II and III, while the typical spatial resolution of silicon strip detectors is between 5 and 10 microns.

In \autoref{fig:Layout} we show a possible paradigmatic example of a ERMES facility (ERMES stands for Earthquake Reconnaissance via Muon beam Evolution in Silicon dioxide). 
In the uppermost diagram, we show a typical scenario of an incipient fault, i.e. a fracture in the Earth’s crust along which blocks of rock move relative to each other.  Here we show, for the sake of simplicity, a single fault line/plane instead of a real scenario consisting of a network of faults typical of a seismically active area. 
The tectonic pressure (red arrows) is acting across the fault line: this is the forerunner driving effect, which has to be monitored. 
Hence, the muon beam must be launched across the fault line/surface at some orthogonal angle and at some adequate depth underground, in such a way that it can be subject to the PWR effect, as depicted in the lowest diagram. 

That implies a location on the ground surface of the accelerator complex devoted to generate the muon beam, which is then launched underground through a well, hosting the muon transport line. 
If necessary, dedicated 90$^\circ$ deflection systems launch the muon perpendicularly across the fault, up to the detector cave, located at nearly the same depth as the deflection system cave. 
A more realistic scenario consists in a switch-yard of several muon beam lines, driven by the same muon accelerator located on the ground surface, which sample the crust across multiple fault lines of the same fault network. 
Each muon beam line is directed to the surface into a dedicated detector housing capable of measuring individual muon beams in a multiplexing way, sharing the muon accelerator repetition rate over all muon beam lines.
When the fault line/plane of interest is located at or near the surface in mountainous regions, the muon facility design can be significantly simplified by directing the muon beam underground quasi-horizontally without necessarily requiring significant deflections (see \autoref{fig:Layout}).

Thanks to the rather modest requirements of the ERMES systems for the muon beam emittance, as well as the muon bunch population, as discussed in previous sections, the muon accelerator complex is expected to deliver modest performances compared to present muon collider scenarios. This opens the possibility of employing plasma-based accelerators and compact muon sources, potentially enabling a ground-based accelerator complex with a footprint comparable to that of a modern synchrotron facility.

Muons production by plasma accelerated electron bunches and use of the Bethe-Heitler process in a high Z-target has been recently demonstrated \cite{calvin2025experimental}. 
The next generation of 100\,J - 100\,Hz repetition rate class, high power lasers, together with newly conceived meter scale plasma targets \cite{rockafellow2025high}, will enable the acceleration of nC electron bunches at 10 GeV energies \cite{picksley2024matched}. 
Following the scalings found in \cite{titov2009dimuon} and Monte Carlo simulations presented in \cite{rao2018bright}, all the main components enabling the production of ultrashort, GeV class muon bunches seem to be within reach in the next decade.

In order to realize a 500\,GeV muon source by Laser WakeField Acceleration (LWFA) many plasma stages need to be employed for overcoming the dephasing between the laser and particles \cite{esarey2009physics} and extend the maximum energy to the desired value. 
A staging proof of principle experiment was presented in \cite{steinke2016multistage} while a facility dedicated to staging and transport optimization, also employing active plasma lenses \cite{van2015active, pompili2018focusing}, was recently proposed \cite{lindstrom2025sparta}.

With all the aforementioned building blocks, we can propose a conceptual layout for a compact, 500\,GeV class muon source. 
The first component would be a plasma based electron accelerator delivering $\lesssim$10\,nC electron bunch at 10\,GeV on a solid, high Z-target of proper thickness. Keeping into account also the muons produced by decaying pions, FLUKA simulations indicate a yield of about $5 \times 10^{-4}$ muons per primary electron, resulting in $\approx 10^7-10^8$ muons per shot. Notice that $\mu^+\,\mu^-$ pairs are produced at this stage; plasma acceleration works properly for negative particles, since the forces are usually defocusing and decelerating for positive ones.

The produced muon bunch would have an energy in excess of 2\,GeV and a divergence of few degrees, together with a short length ($\lesssim$ 1\,ps) mimicking that of the driving electron bunch. 
With an active plasma lens, the negative muons are focused in a 0.5\,m long plasma stage with a plasma density of the order $10^{17}$\,cm$^{-3}$, driven by a 100's TW class high power laser for acceleration, while positive ones would be defocused and lost at this point. The few close to axis will experience decelerating and defocusing plasma fields and would likely decay there. 

At the assumed energy, muons are already relativistic and the bunch length would increase significantly only for geometric effects due to the relatively large divergence ($\lesssim$ 100\,mrad) after production. As detailed in \cite{drobniak2025development}, this problem, and the ones arising from chromaticity in focusing, could be solved by use of carefully tailored non-linear active plasma lenses; due to fast acceleration and beam cooling, those issues are likely to be prominent only in the very first few plasma stages. 
In addition, since the number of betatron oscillations in a single plasma stage is limited and beam quality is not a concern for our target application, requirements for the bunch matched size can be loosened. 

As shown in \cite{rossi2020angstrom} for electrons, the envisioned plasma stages can reach an accelerating field od 10\,GV/m, allowing to increase muons energy of 5\,GeV each with a high charge transmission rate ($\ge 90\%$) that will rapidly approach unity as energy is increased. 
Overall, a $\approx 50$\,m active acceleration length is required; considering the active plasma senses needed for coupling successive plasma stages, an average 1\,m length can be envisioned for each plasma module, totaling an overall $\approx$ 100\,m accelerator length. 
Due to a $q/m$ ratio 200 times smaller than electron's, the muons bunch could be safely U-turned by 10\,T superconducting bending magnets in a $\approx$ 80 m radius (at 250\,GeV or less) or in about half that value by curved discharge capillaries \cite{pompili2024guiding, frazzitta2024theory}, allowing to fold the accelerator and reduce its overall length to dimensions of a modern synchrotron facility.

\section{Proof-of-principle experiments} \label{Sec:PoP}

In this section, we describe two series of proof-of-principle (p.o.p.) experiments aimed at checking the piezoelectric effect of deflecting a charged particle which is crossing quartz crystals excited by piezoelectricity to produce an electric field inside them. 
The first category of p.o.p. experiment adopts a beam of electrons injected into a single quartz crystal subject to an external pressure generated by a bench press. 
Here, the deflection is applied along the known direction of the quartz crystal axis, either towards the positive or negative versus. 
This is a deterministic, well tunable and controllable effect, as described in the first paragraph of this section. 
Since the thickness of a single isolated quartz crystal can be as small as a few millimeters, one can use an electron beam of moderate energy, instead of a muon beam, for the sake of simplicity and easiness in setting up and operating the experiment. 
The second category of p.o.p. experiment, illustrated in the second paragraph of this section, aims at investigating the random walk process applied to a low energy muon beam traversing a short section, a few meter long, of crystalline rock (granite) with embedded a number of quartz crystals, laying on a bench press capable to drive significant pressure on the rock sample, i.e. in the range of a few hundreds of MPa. 
For this class of p.o.p. experiments muons are needed, so to be capable to propagate through meters of granite: the muon energy requested is, as discussed below, in the range of a few GeV. 
These muons are nowadays available at a number of test-facilities world-wide \cite{IPAC-M2}.

\subsection{P.O.P. experiments using single quartz crystals or thin polycrystalline aggregates}
Here we describe how an electron beam can probe the piezoelectric response of a single quartz crystal under controlled applied pressure. 
Modern electron linacs can readily deliver beams in the 20--150\,MeV energy range, and compact high-pressure presses (roughly 1\,m$^3$ in overall size) are available. 
A quartz crystal measuring approximately 0.5--1\,cm per side can be mounted within such a press, positioned just downstream of the linac exit. 
By varying the applied load, one can map the piezoelectric field generated in the crystal up to the point of mechanical failure.

To evaluate the feasibility of such an experiment, we carried out numerical simulations using the well-established tracking code \textsc{ASTRA} \cite{floettmann2011astra}. 
Beam tracking was performed after importing a 3D field map of volume $10 \times 10 \times 10$\,mm$^3$, modeling a uniform piezoelectric field oriented perpendicular to the beam direction. \textsc{ASTRA} also accounts for space-charge effects, which are particularly relevant at lower beam energies.

\begin{figure}[htbp]
    \centering
    \includegraphics[width=0.8\linewidth]{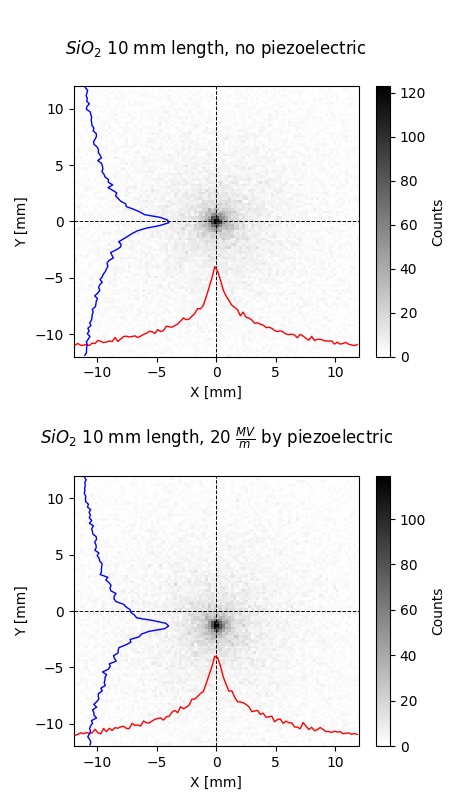}
    \caption{2D histograms of a 150\,MeV electron beam passing through a 10\,mm SiO$_2$ crystal and downstream propagates 1 m. The top image shows multiple Coulomb scattering only, without piezoelectric effects; the bottom image includes both Coulomb scattering and a 20\,MV/m piezoelectric field induced by pressure on the crystal. After the crystal, electrons drift 1\,m before reaching a hypothetical diagnostic target.}
    \label{fig:electrons_1}
\end{figure}

We first considered a 150\,MeV electron beam ($\sigma_x = 200\,\upmu$m, $\varepsilon_n = 1$\,mm$\cdot$mrad) incident perpendicularly on a $10 \times 10 \times 10$\,mm$^3$ quartz crystal. 
First, the beam was propagated through the uniform field within the crystal using \textsc{ASTRA}. After exiting the crystal, MCS effect was applied to account for stochastic angular broadening, as calculated using Molière’s theory (\autoref{eq:moliere}).
The beam then drifts for 1\,m to a downstream diagnostic target. 
In the absence of pressure, the centroid remains on axis (\autoref{fig:electrons_1}, upper plot). 
When a piezoelectric field of 20\,MV/m is applied, the centroid shifts by approximately 1.5\,mm (\autoref{fig:electrons_1}, bottom plot).
Although MCS significantly increases beam emittance and divergence—potentially requiring an iris in the experimental setup—the centroid displacement remains distinct and measurable.

\begin{figure}[htbp]
    \centering
    \includegraphics[width=1\linewidth]{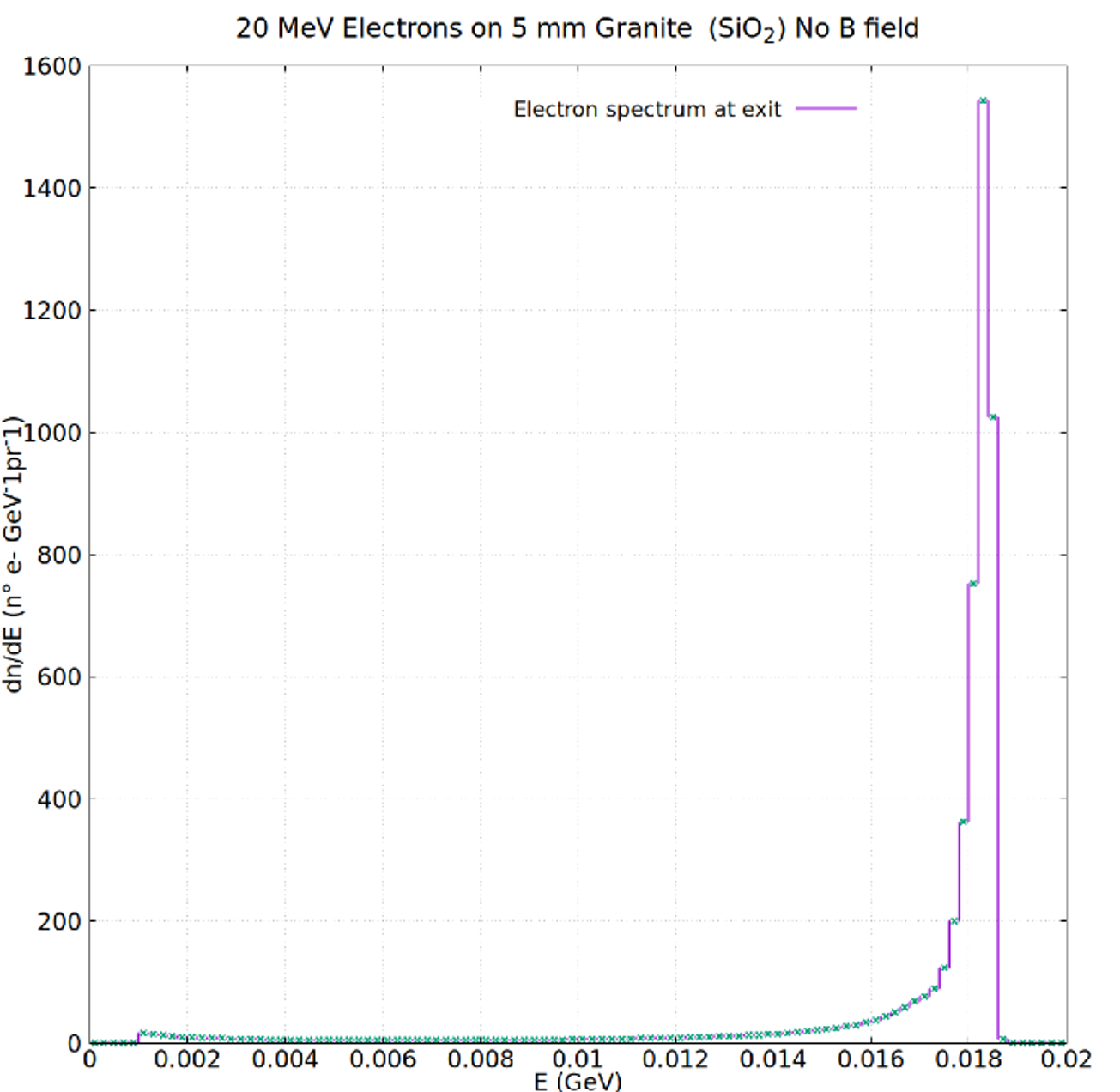}
    \caption{Energy spectrum of 20 MeV electrons crossing a 5 mm thick quartz crystal}
    \label{fig:espectra}
\end{figure}

We also simulated a 20\,MeV beam traversing a shorter crystal ($10 \times 10 \times 5$\,mm$^3$) to limit MCS.
At this lower energy, also considering a low 5\,MV/m piezo field still produces a clear centroid shift of $\sim0.8$\,mm after a 60\,cm drift, for this lower energy we consider a shorter drift to limit the beam enlargement. In order to check the energy loss of 20 MeV electrons in a 5\,mm thick quartz crystal, we ran FLUKA and we found at the crystal exit the energy spectrum shown in \autoref{fig:espectra}. The spectrum peak is centered at about 20\,MeV, with a sharp tail down to 15-16\,MeV, showing that the electron beam crossing the quartz crystal is still very measurable.

Both energy cases -we have already identified laboratories where the experimental test could be performed \cite{giuliano2025compact,valente2006diagnostics} - demonstrate that piezoelectric fields $ \geq 5$\,MV/m produce measurable centroid deflections (\autoref{tab:astra-results}), validating the experimental approach. 
This proof-of-principle study paves the way for an actual experiment to characterize the pressure-to-field relationship and investigate nonlinear behaviors as quartz approaches fracture.

\begin{table}[ht]
\centering
\caption{Summary of beam centroid deflection $\langle y \rangle$ from \textsc{ASTRA} simulations for two different beam energies and crystal configurations, under varying transverse electric field strengths.}
\begin{tabular}{cccccc}
\hline
\textbf{En} & \textbf{Drift} & \textbf{SiO$_2$-length} & \textbf{Piezo \(E_y\)} & \(\langle y \rangle\) \textbf{Deflect.} \\
\textbf{[MeV]}  & \textbf{[m]}  & \textbf{[mm]}           & \textbf{[MV/m]}       & \textbf{[mm]}         \\
\hline
150 & 1  & 10 & 20 & 1.5 \\
150 & 1  & 10 & 5  & 0.4 \\
20  & 0.6   & 5 & 20 & 3.0 \\
20  & 0.6   & 5 & 5  & 0.77 \\
\hline
\end{tabular}
\label{tab:astra-results}
\end{table}

\subsection{P.O.P. experiments using meter thick granite/gneiss slabs}

Proof of principle experiments with GeV muons should be carried out using granite bars in sections of meter-sized length, compatible with available bench presses, with nearly $20\times20$\,cm section, through which muons are propagated to sample the stochastically arranged piezoelectric fields of randomly oriented quartz crystals. In order to maximize the associated PRW effect granite rocks with large quartz crystals should be chosen, as this enhances the key parameter $\Delta$, expressed in \autoref{eq:Delta}. Using 5\,GeV muons crossing 5 meter overall granite length, with 2\,cm quartz crystals ($L_c=0.02$) and 1\,cm neutral gap ($f_c=2/3$), the theoretical model described in \autoref{Sec:Theory} provides following results, if a piezoelectric field $E_p=25$\,MV/m, corresponding to an induced pressure of 5\,kbar (500\,MPa), is assumed: $\sigma_{r^\prime}=1.64$\,mrad and  $\sigma_r=5$\,mm for PRW, and  $\theta_\text{MCS}=40$\,mrad, $\sigma_\text{TOT}=9$\,cm  for MCS. Therefore $\Delta=0.0015$ is expected, implying a beam spot size enhancement of about 140$\,\upmu$m due to PRW over 9\,cm overall beam spot size, just at exit from the granite bars.

The same set of parameters has been tested using MuAEGIS to propagate a beam with an initial round spot size of $\sigma_{r0} = 1$\,cm and normalized emittance $\varepsilon_{n0} = 10^3$\,mm$\cdot$mrad.  
These initial conditions correspond to a $\beta$-function on the order of the slab length, ensuring that the beam remains transversely confined within the slab throughout the propagation.  
The average beam kinetic energy at the slab exit is 2.76\,MeV/$c$.

Accounting for both MCS and PRW effects, the final beam spot size is $\sigma_\text{TOT} = 73.132$\,mm with an angular spread of $\sigma^\prime_\text{TOT} = 30.686$\,mrad.  
When the PRW effect is suppressed---effectively modeling the case where the applied pressure is released---the resulting spot size is $\sigma_\text{MCS} = 73.005$\,mm with $\sigma^\prime_\text{MCS} = 30.626$\,mrad.

The observed enhancement in the transverse beam size due to the PRW is approximately $130\,\upmu$m, corresponding to a relative increase of $\Delta = 1.741 \times 10^{-3}$.

One further mechanism of investigation should be the potential decrease of MCS due to muon channeling effects, as described by \cite{Korotchenko2020}, in the crystals forming the granite slabs. Indeed, if muons are subject to channeling in going through the different types of crystals that form the rock sample (not only quartz but also feldspar and mica), this could significantly affect the scattering behavior.

\section{CONCLUSION} \label{Sec:Conclusion}
We name ERMES the technique described in this paper, an acronym for Earthquake Reconnaissance via Muon beam Evolution in Silicon dioxide. The name also echoes Hermes, the messenger of the gods in Greek mythology, symbolically evoking the role of high-energy muons as swift carriers of information through the Earth’s crust, potentially capable of anticipating large seismic events. 
We also hope that vigorous R\&D will be carried out to address the several challenges posed by the construction and test of an ERMES facility.
In particular, this involves developing a conceptual design and conducting related experimental tests to lay the groundwork for a real system capable of providing early warnings of major earthquakes. The proof of principle experiments are absolutely necessary in order to demonstrate the conceptual ideas described in this paper, beyond the evidence of theoretical analysis and simulative predictions here illustrated.

We would also like to underline two further relevant subjects for future investigations.

The first area of investigation should focus on coherent piezoelectric effects on the muon beam, as extensively discussed in \cite{bishop1981piezoelectric}, where these special conditions are named ``true piezoelectric''. These could lead to a true transverse deflection of the whole muon beam, opposed to a statistical random-walk induced spot size enlargement, as discussed throughout this paper.

The second area of investigation deals with magnetization of rocks that could also lead to either a statistical effects or a coherent steering effects. A stress field acting on a crystalline rock  influences the growth, reorientation, and mechanical concentration of ferromagnetic minerals—such as magnetite—along preferred directions. This process is reflected in a stress-controlled modification of the magnetic susceptibility ellipsoid, whose principal axes tend to align with the principal directions of stress.
The maximum susceptibility axis (Kmax) typically aligns with the extension direction, and the minimum (Kmin) with the compression direction (e.g., \cite{hrouda1982magnetic}).
    
    Pure single-domain magnetite has a high magnetic moment, approximately 92\,Am$^2$/kg, but for larger multi-domain magnetite grains, typical of granitic rocks, the magnetic moment decreases sensibly, and in addition, in such rocks, magnetite is usually an accessory mineral (less than 1\% by volume), and is generally distributed randomly, yielding relatively weak net remanent magnetizations.
However, under sufficient tectonic stress capable of producing foliation, magnetite grains may be mechanically concentrated and aligned along foliation planes, or may chemically grow along such planes from previous phases.
In these conditions, the alignment of magnetic moments becomes more efficient, potentially generating stronger and more anisotropic net remanent magnetizations. According to \cite{OTOFUJI2000271}, the remanent magnetization in foliated granites can be four times higher than in unfoliated ones. This enhanced magnetization may influence the propagation of high-energy muons via the Lorentz force.
Although the piezoelectric contribution from quartz—often 40–50\% by volume—likely dominates the electromagnetic interaction with the muon beam, the role of the contribution of stress-induced magnetization in muon trajectory perturbation remains plausible and merits further investigation.
\vspace{6 mm}

\section{ACKNOWLEDGMENTS}

We are very thankful to Marco Zanetti (Università di Padova) and Paolo Checchia (INFN-Padova) for many useful discussions about muon detectors and their performances. We also thank Luigi Palumbo (Università di Roma La Sapienza) for his strong encouragement.


\begin{thebibliography}{40}%
\makeatletter
\providecommand \@ifxundefined [1]{%
 \@ifx{#1\undefined}
}%
\providecommand \@ifnum [1]{%
 \ifnum #1\expandafter \@firstoftwo
 \else \expandafter \@secondoftwo
 \fi
}%
\providecommand \@ifx [1]{%
 \ifx #1\expandafter \@firstoftwo
 \else \expandafter \@secondoftwo
 \fi
}%
\providecommand \natexlab [1]{#1}%
\providecommand \enquote  [1]{``#1''}%
\providecommand \bibnamefont  [1]{#1}%
\providecommand \bibfnamefont [1]{#1}%
\providecommand \citenamefont [1]{#1}%
\providecommand \href@noop [0]{\@secondoftwo}%
\providecommand \href [0]{\begingroup \@sanitize@url \@href}%
\providecommand \@href[1]{\@@startlink{#1}\@@href}%
\providecommand \@@href[1]{\endgroup#1\@@endlink}%
\providecommand \@sanitize@url [0]{\catcode `\\12\catcode `\$12\catcode `\&12\catcode `\#12\catcode `\^12\catcode `\_12\catcode `\%12\relax}%
\providecommand \@@startlink[1]{}%
\providecommand \@@endlink[0]{}%
\providecommand \url  [0]{\begingroup\@sanitize@url \@url }%
\providecommand \@url [1]{\endgroup\@href {#1}{\urlprefix }}%
\providecommand \urlprefix  [0]{URL }%
\providecommand \Eprint [0]{\href }%
\providecommand \doibase [0]{https://doi.org/}%
\providecommand \selectlanguage [0]{\@gobble}%
\providecommand \bibinfo  [0]{\@secondoftwo}%
\providecommand \bibfield  [0]{\@secondoftwo}%
\providecommand \translation [1]{[#1]}%
\providecommand \BibitemOpen [0]{}%
\providecommand \bibitemStop [0]{}%
\providecommand \bibitemNoStop [0]{.\EOS\space}%
\providecommand \EOS [0]{\spacefactor3000\relax}%
\providecommand \BibitemShut  [1]{\csname bibitem#1\endcsname}%
\let\auto@bib@innerbib\@empty
\bibitem [{\citenamefont {Rovida}\ \emph {et~al.}(2022)\citenamefont {Rovida}, \citenamefont {Locati}, \citenamefont {Camassi}, \citenamefont {Lolli}, \citenamefont {Gasperini},\ and\ \citenamefont {Antonucci}}]{RovidaEtAl2022}%
  \BibitemOpen
  \bibfield  {author} {\bibinfo {author} {\bibfnamefont {A.}~\bibnamefont {Rovida}}, \bibinfo {author} {\bibfnamefont {M.}~\bibnamefont {Locati}}, \bibinfo {author} {\bibfnamefont {R.}~\bibnamefont {Camassi}}, \bibinfo {author} {\bibfnamefont {B.}~\bibnamefont {Lolli}}, \bibinfo {author} {\bibfnamefont {P.}~\bibnamefont {Gasperini}},\ and\ \bibinfo {author} {\bibfnamefont {A.}~\bibnamefont {Antonucci}},\ }\href {https://doi.org/10.13127/CPTI/CPTI15.4} {\bibinfo {title} {Catalogo parametrico dei terremoti italiani (cpti15), versione 4.0}} (\bibinfo {year} {2022})\BibitemShut {NoStop}%
\bibitem [{\citenamefont {Navas}\ \emph {et~al.}(2024)\citenamefont {Navas} \emph {et~al.}}]{ParticleDataGroup:2024cfk}%
  \BibitemOpen
  \bibfield  {author} {\bibinfo {author} {\bibfnamefont {S.}~\bibnamefont {Navas}} \emph {et~al.} (\bibinfo {collaboration} {Particle Data Group}),\ }\href {https://doi.org/10.1103/PhysRevD.110.030001} {\bibfield  {journal} {\bibinfo  {journal} {Phys. Rev. D}\ }\textbf {\bibinfo {volume} {110}},\ \bibinfo {pages} {030001} (\bibinfo {year} {2024})}\BibitemShut {NoStop}%
\bibitem [{\citenamefont {IMCC}\ and\ \citenamefont {Mucol}(2025)}]{collider}%
  \BibitemOpen
  \bibfield  {author} {\bibinfo {author} {\bibnamefont {IMCC}}\ and\ \bibinfo {author} {\bibnamefont {Mucol}},\ }\href {https://muoncollider.web.cern.ch/events/imcc-and-mucol-annual-meeting-2025} {\bibfield  {journal} {\bibinfo  {journal} {Desy}\ } (\bibinfo {year} {2025})}\BibitemShut {NoStop}%
\bibitem [{\citenamefont {Ahdida}\ \emph {et~al.}(2022)\citenamefont {Ahdida}, \citenamefont {Bozzato}, \citenamefont {Calzolari}, \citenamefont {Cerutti}, \citenamefont {Charitonidis}, \citenamefont {Cimmino}, \citenamefont {Coronetti}, \citenamefont {D’Alessandro}, \citenamefont {Servelle}, \citenamefont {Esposito}, \citenamefont {Froeschl}, \citenamefont {Alía}, \citenamefont {Gerbershagen}, \citenamefont {Gilardoni}, \citenamefont {Horváth}, \citenamefont {Hugo}, \citenamefont {Infantino}, \citenamefont {Kouskoura}, \citenamefont {Lechner}, \citenamefont {Lefebvre}, \citenamefont {Lerner}, \citenamefont {Magistris}, \citenamefont {Manousos}, \citenamefont {Moryc}, \citenamefont {Ruiz}, \citenamefont {Pozzi}, \citenamefont {Prelipcean}, \citenamefont {Roesler}, \citenamefont {Rossi}, \citenamefont {Gilarte}, \citenamefont {Pujol}, \citenamefont {Schoofs}, \citenamefont {Stránský}, \citenamefont {Theis}, \citenamefont {Tsinganis}, \citenamefont {Versaci}, \citenamefont {Vlachoudis}, \citenamefont
  {Waets},\ and\ \citenamefont {Widorski}}]{Ahdida2022FLUKA}%
  \BibitemOpen
  \bibfield  {author} {\bibinfo {author} {\bibfnamefont {C.}~\bibnamefont {Ahdida}}, \bibinfo {author} {\bibfnamefont {D.}~\bibnamefont {Bozzato}}, \bibinfo {author} {\bibfnamefont {D.}~\bibnamefont {Calzolari}}, \bibinfo {author} {\bibfnamefont {F.}~\bibnamefont {Cerutti}}, \bibinfo {author} {\bibfnamefont {N.}~\bibnamefont {Charitonidis}}, \bibinfo {author} {\bibfnamefont {A.}~\bibnamefont {Cimmino}}, \bibinfo {author} {\bibfnamefont {A.}~\bibnamefont {Coronetti}}, \bibinfo {author} {\bibfnamefont {G.~L.}\ \bibnamefont {D’Alessandro}}, \bibinfo {author} {\bibfnamefont {A.~D.}\ \bibnamefont {Servelle}}, \bibinfo {author} {\bibfnamefont {L.~S.}\ \bibnamefont {Esposito}}, \bibinfo {author} {\bibfnamefont {R.}~\bibnamefont {Froeschl}}, \bibinfo {author} {\bibfnamefont {R.~G.}\ \bibnamefont {Alía}}, \bibinfo {author} {\bibfnamefont {A.}~\bibnamefont {Gerbershagen}}, \bibinfo {author} {\bibfnamefont {S.}~\bibnamefont {Gilardoni}}, \bibinfo {author} {\bibfnamefont {D.}~\bibnamefont {Horváth}}, \bibinfo
  {author} {\bibfnamefont {G.}~\bibnamefont {Hugo}}, \bibinfo {author} {\bibfnamefont {A.}~\bibnamefont {Infantino}}, \bibinfo {author} {\bibfnamefont {V.}~\bibnamefont {Kouskoura}}, \bibinfo {author} {\bibfnamefont {A.}~\bibnamefont {Lechner}}, \bibinfo {author} {\bibfnamefont {B.}~\bibnamefont {Lefebvre}}, \bibinfo {author} {\bibfnamefont {G.}~\bibnamefont {Lerner}}, \bibinfo {author} {\bibfnamefont {M.}~\bibnamefont {Magistris}}, \bibinfo {author} {\bibfnamefont {A.}~\bibnamefont {Manousos}}, \bibinfo {author} {\bibfnamefont {G.}~\bibnamefont {Moryc}}, \bibinfo {author} {\bibfnamefont {F.~O.}\ \bibnamefont {Ruiz}}, \bibinfo {author} {\bibfnamefont {F.}~\bibnamefont {Pozzi}}, \bibinfo {author} {\bibfnamefont {D.}~\bibnamefont {Prelipcean}}, \bibinfo {author} {\bibfnamefont {S.}~\bibnamefont {Roesler}}, \bibinfo {author} {\bibfnamefont {R.}~\bibnamefont {Rossi}}, \bibinfo {author} {\bibfnamefont {M.~S.}\ \bibnamefont {Gilarte}}, \bibinfo {author} {\bibfnamefont {F.~S.}\ \bibnamefont {Pujol}}, \bibinfo
  {author} {\bibfnamefont {P.}~\bibnamefont {Schoofs}}, \bibinfo {author} {\bibfnamefont {V.}~\bibnamefont {Stránský}}, \bibinfo {author} {\bibfnamefont {C.}~\bibnamefont {Theis}}, \bibinfo {author} {\bibfnamefont {A.}~\bibnamefont {Tsinganis}}, \bibinfo {author} {\bibfnamefont {R.}~\bibnamefont {Versaci}}, \bibinfo {author} {\bibfnamefont {V.}~\bibnamefont {Vlachoudis}}, \bibinfo {author} {\bibfnamefont {A.}~\bibnamefont {Waets}},\ and\ \bibinfo {author} {\bibfnamefont {M.}~\bibnamefont {Widorski}},\ }\href {https://doi.org/10.3389/fphy.2021.788253} {\bibfield  {journal} {\bibinfo  {journal} {Frontiers in Physics}\ }\textbf {\bibinfo {volume} {9}},\ \bibinfo {pages} {788253} (\bibinfo {year} {2022})}\BibitemShut {NoStop}%
\bibitem [{\citenamefont {Bishop}(1981)}]{bishop1981piezoelectric}%
  \BibitemOpen
  \bibfield  {author} {\bibinfo {author} {\bibfnamefont {J.~R.}\ \bibnamefont {Bishop}},\ }\href {https://doi.org/10.1016/0040-1951(81)90268-7} {\bibfield  {journal} {\bibinfo  {journal} {Tectonophysics}\ }\textbf {\bibinfo {volume} {77}},\ \bibinfo {pages} {297} (\bibinfo {year} {1981})}\BibitemShut {NoStop}%
\bibitem [{\citenamefont {Byerlee}(1967)}]{Byerlee1967}%
  \BibitemOpen
  \bibfield  {author} {\bibinfo {author} {\bibfnamefont {J.~D.}\ \bibnamefont {Byerlee}},\ }\href {https://doi.org/10.1063/1.1710026} {\bibfield  {journal} {\bibinfo  {journal} {Journal of Applied Physics}\ }\textbf {\bibinfo {volume} {38}},\ \bibinfo {pages} {2928} (\bibinfo {year} {1967})}\BibitemShut {NoStop}%
\bibitem [{\citenamefont {Byerlee}(1968)}]{Byerlee1968}%
  \BibitemOpen
  \bibfield  {author} {\bibinfo {author} {\bibfnamefont {J.~D.}\ \bibnamefont {Byerlee}},\ }\href {https://doi.org/10.1029/JB073i014p04741} {\bibfield  {journal} {\bibinfo  {journal} {Journal of Geophysical Research}\ }\textbf {\bibinfo {volume} {73}},\ \bibinfo {pages} {4741} (\bibinfo {year} {1968})}\BibitemShut {NoStop}%
\bibitem [{\citenamefont {ichiro Karato}(2008)}]{Karato2008}%
  \BibitemOpen
  \bibfield  {author} {\bibinfo {author} {\bibfnamefont {S.}~\bibnamefont {ichiro Karato}},\ }\href@noop {} {\emph {\bibinfo {title} {Deformation of Earth Materials: An Introduction to the Rheology of Solid Earth}}}\ (\bibinfo  {publisher} {Cambridge University Press},\ \bibinfo {address} {Cambridge, UK},\ \bibinfo {year} {2008})\BibitemShut {NoStop}%
\bibitem [{\citenamefont {Ramsay}(1967)}]{Ramsay1967}%
  \BibitemOpen
  \bibfield  {author} {\bibinfo {author} {\bibfnamefont {J.~G.}\ \bibnamefont {Ramsay}},\ }\href@noop {} {\emph {\bibinfo {title} {Folding and Fracturing of Rocks}}}\ (\bibinfo  {publisher} {McGraw-Hill},\ \bibinfo {address} {New York},\ \bibinfo {year} {1967})\BibitemShut {NoStop}%
\bibitem [{\citenamefont {Ranalli}(1987)}]{Ranalli1995}%
  \BibitemOpen
  \bibfield  {author} {\bibinfo {author} {\bibfnamefont {G.}~\bibnamefont {Ranalli}},\ }\href@noop {} {\emph {\bibinfo {title} {Rheology of the Earth : Deformation and Flow Processes in Geophysics and Geodynamics}}}\ (\bibinfo  {publisher} {Allen \& Unwin},\ \bibinfo {address} {Boston},\ \bibinfo {year} {1987})\BibitemShut {NoStop}%
\bibitem [{\citenamefont {Liu}\ \emph {et~al.}(2023)\citenamefont {Liu}, \citenamefont {Wang}, \citenamefont {Li}, \citenamefont {Wang},\ and\ \citenamefont {Selvadurai}}]{LiuEtAl2023}%
  \BibitemOpen
  \bibfield  {author} {\bibinfo {author} {\bibfnamefont {Z.}~\bibnamefont {Liu}}, \bibinfo {author} {\bibfnamefont {H.}~\bibnamefont {Wang}}, \bibinfo {author} {\bibfnamefont {Y.}~\bibnamefont {Li}}, \bibinfo {author} {\bibfnamefont {X.}~\bibnamefont {Wang}},\ and\ \bibinfo {author} {\bibfnamefont {A.~P.~S.}\ \bibnamefont {Selvadurai}},\ }\href {https://doi.org/10.1007/s00603-022-03095-1} {\bibfield  {journal} {\bibinfo  {journal} {Rock Mechanics and Rock Engineering}\ }\textbf {\bibinfo {volume} {56}},\ \bibinfo {pages} {911} (\bibinfo {year} {2023})}\BibitemShut {NoStop}%
\bibitem [{\citenamefont {Yang}\ \emph {et~al.}(2013)\citenamefont {Yang}, \citenamefont {Xu}, \citenamefont {Wang}, \citenamefont {Nie},\ and\ \citenamefont {Ren}}]{YangEtAl2013}%
  \BibitemOpen
  \bibfield  {author} {\bibinfo {author} {\bibfnamefont {H.~T.}\ \bibnamefont {Yang}}, \bibinfo {author} {\bibfnamefont {J.}~\bibnamefont {Xu}}, \bibinfo {author} {\bibfnamefont {L.}~\bibnamefont {Wang}}, \bibinfo {author} {\bibfnamefont {M.}~\bibnamefont {Nie}},\ and\ \bibinfo {author} {\bibfnamefont {H.~N.}\ \bibnamefont {Ren}},\ }\href@noop {} {\bibfield  {journal} {\bibinfo  {journal} {Chinese Journal of Underground Space and Engineering}\ }\textbf {\bibinfo {volume} {9}},\ \bibinfo {pages} {96} (\bibinfo {year} {2013})}\BibitemShut {NoStop}%
\bibitem [{\citenamefont {Paterson}\ and\ \citenamefont {Wong}(2005)}]{PatersonWong2005}%
  \BibitemOpen
  \bibfield  {author} {\bibinfo {author} {\bibfnamefont {M.~S.}\ \bibnamefont {Paterson}}\ and\ \bibinfo {author} {\bibfnamefont {T.-F.}\ \bibnamefont {Wong}},\ }\href@noop {} {\emph {\bibinfo {title} {Experimental Rock Deformation -- The Brittle Field}}}\ (\bibinfo  {publisher} {Springer-Verlag},\ \bibinfo {address} {Berlin},\ \bibinfo {year} {2005})\BibitemShut {NoStop}%
\bibitem [{\citenamefont {Barrett}\ \emph {et~al.}(1952)\citenamefont {Barrett}, \citenamefont {Bollinger}, \citenamefont {Cocconi}, \citenamefont {Eisenberg},\ and\ \citenamefont {Greisen}}]{dEdX}%
  \BibitemOpen
  \bibfield  {author} {\bibinfo {author} {\bibfnamefont {P.~H.}\ \bibnamefont {Barrett}}, \bibinfo {author} {\bibfnamefont {L.~M.}\ \bibnamefont {Bollinger}}, \bibinfo {author} {\bibfnamefont {G.}~\bibnamefont {Cocconi}}, \bibinfo {author} {\bibfnamefont {Y.}~\bibnamefont {Eisenberg}},\ and\ \bibinfo {author} {\bibfnamefont {K.}~\bibnamefont {Greisen}},\ }\href {https://doi.org/10.1103/RevModPhys.24.133} {\bibfield  {journal} {\bibinfo  {journal} {Rev. Mod. Phys.}\ }\textbf {\bibinfo {volume} {24}},\ \bibinfo {pages} {133} (\bibinfo {year} {1952})}\BibitemShut {NoStop}%
\bibitem [{\citenamefont {Molière}(1947)}]{Moliere47}%
  \BibitemOpen
  \bibfield  {author} {\bibinfo {author} {\bibfnamefont {G.}~\bibnamefont {Molière}},\ }\href@noop {} {\bibfield  {journal} {\bibinfo  {journal} {Z. Naturforsch.}\ }\textbf {\bibinfo {volume} {2A}},\ \bibinfo {pages} {133} (\bibinfo {year} {1947})}\BibitemShut {NoStop}%
\bibitem [{\citenamefont {Borozdin}\ \emph {et~al.}(2003)\citenamefont {Borozdin}, \citenamefont {Hogan}, \citenamefont {Morris} \emph {et~al.}}]{Borozdin2003}%
  \BibitemOpen
  \bibfield  {author} {\bibinfo {author} {\bibfnamefont {K.}~\bibnamefont {Borozdin}}, \bibinfo {author} {\bibfnamefont {G.}~\bibnamefont {Hogan}}, \bibinfo {author} {\bibfnamefont {C.}~\bibnamefont {Morris}}, \emph {et~al.},\ }\href {https://doi.org/10.1038/422277a} {\bibfield  {journal} {\bibinfo  {journal} {Nature}\ }\textbf {\bibinfo {volume} {422}},\ \bibinfo {pages} {277} (\bibinfo {year} {2003})},\ \bibinfo {note} {issue Date: 20 March 2003}\BibitemShut {NoStop}%
\bibitem [{\citenamefont {Groom}\ \emph {et~al.}(2001)\citenamefont {Groom}, \citenamefont {Mokhov},\ and\ \citenamefont {Striganov}}]{Groom2001Muon}%
  \BibitemOpen
  \bibfield  {author} {\bibinfo {author} {\bibfnamefont {D.~E.}\ \bibnamefont {Groom}}, \bibinfo {author} {\bibfnamefont {N.~V.}\ \bibnamefont {Mokhov}},\ and\ \bibinfo {author} {\bibfnamefont {S.~I.}\ \bibnamefont {Striganov}},\ }\href {https://doi.org/10.1006/adnd.2001.0861} {\bibfield  {journal} {\bibinfo  {journal} {Atomic Data and Nuclear Data Tables}\ }\textbf {\bibinfo {volume} {78}},\ \bibinfo {pages} {183} (\bibinfo {year} {2001})}\BibitemShut {NoStop}%
\bibitem [{\citenamefont {Harris}\ \emph {et~al.}(2020)\citenamefont {Harris}, \citenamefont {Millman}, \citenamefont {van~der Walt}, \citenamefont {Gommers}, \citenamefont {Virtanen}, \citenamefont {Cournapeau}, \citenamefont {Wieser}, \citenamefont {Taylor}, \citenamefont {Berg}, \citenamefont {Smith}, \citenamefont {Kern}, \citenamefont {Picus}, \citenamefont {Hoyer}, \citenamefont {van Kerkwijk}, \citenamefont {Brett}, \citenamefont {Haldane}, \citenamefont {del R{\'i}o}, \citenamefont {Wiebe}, \citenamefont {Peterson}, \citenamefont {G{\'e}rard-Marchant}, \citenamefont {Sheppard}, \citenamefont {Reddy}, \citenamefont {Weckesser}, \citenamefont {Abbasi}, \citenamefont {Gohlke},\ and\ \citenamefont {Oliphant}}]{harris2020array}%
  \BibitemOpen
  \bibfield  {author} {\bibinfo {author} {\bibfnamefont {C.~R.}\ \bibnamefont {Harris}}, \bibinfo {author} {\bibfnamefont {K.~J.}\ \bibnamefont {Millman}}, \bibinfo {author} {\bibfnamefont {S.~J.}\ \bibnamefont {van~der Walt}}, \bibinfo {author} {\bibfnamefont {R.}~\bibnamefont {Gommers}}, \bibinfo {author} {\bibfnamefont {P.}~\bibnamefont {Virtanen}}, \bibinfo {author} {\bibfnamefont {D.}~\bibnamefont {Cournapeau}}, \bibinfo {author} {\bibfnamefont {E.}~\bibnamefont {Wieser}}, \bibinfo {author} {\bibfnamefont {J.}~\bibnamefont {Taylor}}, \bibinfo {author} {\bibfnamefont {S.}~\bibnamefont {Berg}}, \bibinfo {author} {\bibfnamefont {N.~J.}\ \bibnamefont {Smith}}, \bibinfo {author} {\bibfnamefont {R.}~\bibnamefont {Kern}}, \bibinfo {author} {\bibfnamefont {M.}~\bibnamefont {Picus}}, \bibinfo {author} {\bibfnamefont {S.}~\bibnamefont {Hoyer}}, \bibinfo {author} {\bibfnamefont {M.~H.}\ \bibnamefont {van Kerkwijk}}, \bibinfo {author} {\bibfnamefont {M.}~\bibnamefont {Brett}}, \bibinfo {author} {\bibfnamefont
  {A.}~\bibnamefont {Haldane}}, \bibinfo {author} {\bibfnamefont {J.~F.}\ \bibnamefont {del R{\'i}o}}, \bibinfo {author} {\bibfnamefont {M.}~\bibnamefont {Wiebe}}, \bibinfo {author} {\bibfnamefont {P.}~\bibnamefont {Peterson}}, \bibinfo {author} {\bibfnamefont {P.}~\bibnamefont {G{\'e}rard-Marchant}}, \bibinfo {author} {\bibfnamefont {K.}~\bibnamefont {Sheppard}}, \bibinfo {author} {\bibfnamefont {T.}~\bibnamefont {Reddy}}, \bibinfo {author} {\bibfnamefont {W.}~\bibnamefont {Weckesser}}, \bibinfo {author} {\bibfnamefont {H.}~\bibnamefont {Abbasi}}, \bibinfo {author} {\bibfnamefont {C.}~\bibnamefont {Gohlke}},\ and\ \bibinfo {author} {\bibfnamefont {T.~E.}\ \bibnamefont {Oliphant}},\ }\href {https://doi.org/10.1038/s41586-020-2649-2} {\bibfield  {journal} {\bibinfo  {journal} {Nature}\ }\textbf {\bibinfo {volume} {585}},\ \bibinfo {pages} {357} (\bibinfo {year} {2020})}\BibitemShut {NoStop}%
\bibitem [{\citenamefont {Abbiendi}\ \emph {et~al.}(2017)\citenamefont {Abbiendi}, \citenamefont {Carloni~Calame}, \citenamefont {Marconi}, \citenamefont {Matteuzzi}, \citenamefont {Montagna}, \citenamefont {Nicrosini}, \citenamefont {Passera}, \citenamefont {Piccinini}, \citenamefont {Tenchini}, \citenamefont {Trentadue},\ and\ \citenamefont {Venanzoni}}]{Abbiendi:2017}%
  \BibitemOpen
  \bibfield  {author} {\bibinfo {author} {\bibfnamefont {G.}~\bibnamefont {Abbiendi}}, \bibinfo {author} {\bibfnamefont {C.~M.}\ \bibnamefont {Carloni~Calame}}, \bibinfo {author} {\bibfnamefont {U.}~\bibnamefont {Marconi}}, \bibinfo {author} {\bibfnamefont {C.}~\bibnamefont {Matteuzzi}}, \bibinfo {author} {\bibfnamefont {G.}~\bibnamefont {Montagna}}, \bibinfo {author} {\bibfnamefont {O.}~\bibnamefont {Nicrosini}}, \bibinfo {author} {\bibfnamefont {M.}~\bibnamefont {Passera}}, \bibinfo {author} {\bibfnamefont {F.}~\bibnamefont {Piccinini}}, \bibinfo {author} {\bibfnamefont {R.}~\bibnamefont {Tenchini}}, \bibinfo {author} {\bibfnamefont {L.}~\bibnamefont {Trentadue}},\ and\ \bibinfo {author} {\bibfnamefont {G.}~\bibnamefont {Venanzoni}},\ }\bibfield  {journal} {\bibinfo  {journal} {Eur.\ Phys.\ J.\ C}\ }\textbf {\bibinfo {volume} {77}},\ \href {https://doi.org/10.1140/epjc/s10052-017-4633-z} {10.1140/epjc/s10052-017-4633-z} (\bibinfo {year} {2017}),\ \bibinfo {note} {published online: 1 March 2017; Received: 17
  October 2016; Accepted: 17 January 2017}\BibitemShut {NoStop}%
\bibitem [{\citenamefont {Calvin}\ \emph {et~al.}(2025)\citenamefont {Calvin}, \citenamefont {Gerstmayr}, \citenamefont {Arran}, \citenamefont {Tudor}, \citenamefont {Foster}, \citenamefont {Bergmann}, \citenamefont {Doria}, \citenamefont {Kettle}, \citenamefont {Maguire}, \citenamefont {Malka} \emph {et~al.}}]{calvin2025experimental}%
  \BibitemOpen
  \bibfield  {author} {\bibinfo {author} {\bibfnamefont {L.}~\bibnamefont {Calvin}}, \bibinfo {author} {\bibfnamefont {E.}~\bibnamefont {Gerstmayr}}, \bibinfo {author} {\bibfnamefont {C.}~\bibnamefont {Arran}}, \bibinfo {author} {\bibfnamefont {L.}~\bibnamefont {Tudor}}, \bibinfo {author} {\bibfnamefont {T.}~\bibnamefont {Foster}}, \bibinfo {author} {\bibfnamefont {B.}~\bibnamefont {Bergmann}}, \bibinfo {author} {\bibfnamefont {D.}~\bibnamefont {Doria}}, \bibinfo {author} {\bibfnamefont {B.}~\bibnamefont {Kettle}}, \bibinfo {author} {\bibfnamefont {H.}~\bibnamefont {Maguire}}, \bibinfo {author} {\bibfnamefont {V.}~\bibnamefont {Malka}}, \emph {et~al.},\ }\href@noop {} {\bibfield  {journal} {\bibinfo  {journal} {arXiv preprint arXiv:2503.20904}\ } (\bibinfo {year} {2025})}\BibitemShut {NoStop}%
\bibitem [{\citenamefont {Rockafellow}\ \emph {et~al.}(2025)\citenamefont {Rockafellow}, \citenamefont {Shrock}, \citenamefont {Miao}, \citenamefont {Sloss}, \citenamefont {Le}, \citenamefont {Hancock}, \citenamefont {Zahedpour}, \citenamefont {Hollinger}, \citenamefont {Wang}, \citenamefont {King} \emph {et~al.}}]{rockafellow2025high}%
  \BibitemOpen
  \bibfield  {author} {\bibinfo {author} {\bibfnamefont {E.}~\bibnamefont {Rockafellow}}, \bibinfo {author} {\bibfnamefont {J.~E.}\ \bibnamefont {Shrock}}, \bibinfo {author} {\bibfnamefont {B.}~\bibnamefont {Miao}}, \bibinfo {author} {\bibfnamefont {A.}~\bibnamefont {Sloss}}, \bibinfo {author} {\bibfnamefont {M.~S.}\ \bibnamefont {Le}}, \bibinfo {author} {\bibfnamefont {S.~W.}\ \bibnamefont {Hancock}}, \bibinfo {author} {\bibfnamefont {S.}~\bibnamefont {Zahedpour}}, \bibinfo {author} {\bibfnamefont {R.~C.}\ \bibnamefont {Hollinger}}, \bibinfo {author} {\bibfnamefont {S.}~\bibnamefont {Wang}}, \bibinfo {author} {\bibfnamefont {J.}~\bibnamefont {King}}, \emph {et~al.},\ }\href@noop {} {\bibfield  {journal} {\bibinfo  {journal} {Nuclear Instruments and Methods in Physics Research Section A: Accelerators, Spectrometers, Detectors and Associated Equipment}\ ,\ \bibinfo {pages} {170586}} (\bibinfo {year} {2025})}\BibitemShut {NoStop}%
\bibitem [{\citenamefont {Picksley}\ \emph {et~al.}(2024)\citenamefont {Picksley}, \citenamefont {Stackhouse}, \citenamefont {Benedetti}, \citenamefont {Nakamura}, \citenamefont {Tsai}, \citenamefont {Li}, \citenamefont {Miao}, \citenamefont {Shrock}, \citenamefont {Rockafellow}, \citenamefont {Milchberg} \emph {et~al.}}]{picksley2024matched}%
  \BibitemOpen
  \bibfield  {author} {\bibinfo {author} {\bibfnamefont {A.}~\bibnamefont {Picksley}}, \bibinfo {author} {\bibfnamefont {J.}~\bibnamefont {Stackhouse}}, \bibinfo {author} {\bibfnamefont {C.}~\bibnamefont {Benedetti}}, \bibinfo {author} {\bibfnamefont {K.}~\bibnamefont {Nakamura}}, \bibinfo {author} {\bibfnamefont {H.}~\bibnamefont {Tsai}}, \bibinfo {author} {\bibfnamefont {R.}~\bibnamefont {Li}}, \bibinfo {author} {\bibfnamefont {B.}~\bibnamefont {Miao}}, \bibinfo {author} {\bibfnamefont {J.}~\bibnamefont {Shrock}}, \bibinfo {author} {\bibfnamefont {E.}~\bibnamefont {Rockafellow}}, \bibinfo {author} {\bibfnamefont {H.}~\bibnamefont {Milchberg}}, \emph {et~al.},\ }\href@noop {} {\bibfield  {journal} {\bibinfo  {journal} {Physical Review Letters}\ }\textbf {\bibinfo {volume} {133}},\ \bibinfo {pages} {255001} (\bibinfo {year} {2024})}\BibitemShut {NoStop}%
\bibitem [{\citenamefont {Titov}\ \emph {et~al.}(2009)\citenamefont {Titov}, \citenamefont {K{\"a}mpfer},\ and\ \citenamefont {Takabe}}]{titov2009dimuon}%
  \BibitemOpen
  \bibfield  {author} {\bibinfo {author} {\bibfnamefont {A.}~\bibnamefont {Titov}}, \bibinfo {author} {\bibfnamefont {B.}~\bibnamefont {K{\"a}mpfer}},\ and\ \bibinfo {author} {\bibfnamefont {H.}~\bibnamefont {Takabe}},\ }\href@noop {} {\bibfield  {journal} {\bibinfo  {journal} {Physical Review Special Topics—Accelerators and Beams}\ }\textbf {\bibinfo {volume} {12}},\ \bibinfo {pages} {111301} (\bibinfo {year} {2009})}\BibitemShut {NoStop}%
\bibitem [{\citenamefont {Rao}\ \emph {et~al.}(2018)\citenamefont {Rao}, \citenamefont {Jeon}, \citenamefont {Kim},\ and\ \citenamefont {Nam}}]{rao2018bright}%
  \BibitemOpen
  \bibfield  {author} {\bibinfo {author} {\bibfnamefont {B.~S.}\ \bibnamefont {Rao}}, \bibinfo {author} {\bibfnamefont {J.~H.}\ \bibnamefont {Jeon}}, \bibinfo {author} {\bibfnamefont {H.~T.}\ \bibnamefont {Kim}},\ and\ \bibinfo {author} {\bibfnamefont {C.~H.}\ \bibnamefont {Nam}},\ }\href@noop {} {\bibfield  {journal} {\bibinfo  {journal} {Plasma Physics and Controlled Fusion}\ }\textbf {\bibinfo {volume} {60}},\ \bibinfo {pages} {095002} (\bibinfo {year} {2018})}\BibitemShut {NoStop}%
\bibitem [{\citenamefont {Esarey}\ \emph {et~al.}(2009)\citenamefont {Esarey}, \citenamefont {Schroeder},\ and\ \citenamefont {Leemans}}]{esarey2009physics}%
  \BibitemOpen
  \bibfield  {author} {\bibinfo {author} {\bibfnamefont {E.}~\bibnamefont {Esarey}}, \bibinfo {author} {\bibfnamefont {C.~B.}\ \bibnamefont {Schroeder}},\ and\ \bibinfo {author} {\bibfnamefont {W.~P.}\ \bibnamefont {Leemans}},\ }\href@noop {} {\bibfield  {journal} {\bibinfo  {journal} {Reviews of modern physics}\ }\textbf {\bibinfo {volume} {81}},\ \bibinfo {pages} {1229} (\bibinfo {year} {2009})}\BibitemShut {NoStop}%
\bibitem [{\citenamefont {Steinke}\ \emph {et~al.}(2016)\citenamefont {Steinke}, \citenamefont {Van~Tilborg}, \citenamefont {Benedetti}, \citenamefont {Geddes}, \citenamefont {Schroeder}, \citenamefont {Daniels}, \citenamefont {Swanson}, \citenamefont {Gonsalves}, \citenamefont {Nakamura}, \citenamefont {Matlis} \emph {et~al.}}]{steinke2016multistage}%
  \BibitemOpen
  \bibfield  {author} {\bibinfo {author} {\bibfnamefont {S.}~\bibnamefont {Steinke}}, \bibinfo {author} {\bibfnamefont {J.}~\bibnamefont {Van~Tilborg}}, \bibinfo {author} {\bibfnamefont {C.}~\bibnamefont {Benedetti}}, \bibinfo {author} {\bibfnamefont {C.}~\bibnamefont {Geddes}}, \bibinfo {author} {\bibfnamefont {C.}~\bibnamefont {Schroeder}}, \bibinfo {author} {\bibfnamefont {J.}~\bibnamefont {Daniels}}, \bibinfo {author} {\bibfnamefont {K.}~\bibnamefont {Swanson}}, \bibinfo {author} {\bibfnamefont {A.}~\bibnamefont {Gonsalves}}, \bibinfo {author} {\bibfnamefont {K.}~\bibnamefont {Nakamura}}, \bibinfo {author} {\bibfnamefont {N.}~\bibnamefont {Matlis}}, \emph {et~al.},\ }\href@noop {} {\bibfield  {journal} {\bibinfo  {journal} {Nature}\ }\textbf {\bibinfo {volume} {530}},\ \bibinfo {pages} {190} (\bibinfo {year} {2016})}\BibitemShut {NoStop}%
\bibitem [{\citenamefont {Van~Tilborg}\ \emph {et~al.}(2015)\citenamefont {Van~Tilborg}, \citenamefont {Steinke}, \citenamefont {Geddes}, \citenamefont {Matlis}, \citenamefont {Shaw}, \citenamefont {Gonsalves}, \citenamefont {Huijts}, \citenamefont {Nakamura}, \citenamefont {Daniels}, \citenamefont {Schroeder} \emph {et~al.}}]{van2015active}%
  \BibitemOpen
  \bibfield  {author} {\bibinfo {author} {\bibfnamefont {J.}~\bibnamefont {Van~Tilborg}}, \bibinfo {author} {\bibfnamefont {S.}~\bibnamefont {Steinke}}, \bibinfo {author} {\bibfnamefont {C.}~\bibnamefont {Geddes}}, \bibinfo {author} {\bibfnamefont {N.}~\bibnamefont {Matlis}}, \bibinfo {author} {\bibfnamefont {B.}~\bibnamefont {Shaw}}, \bibinfo {author} {\bibfnamefont {A.}~\bibnamefont {Gonsalves}}, \bibinfo {author} {\bibfnamefont {J.}~\bibnamefont {Huijts}}, \bibinfo {author} {\bibfnamefont {K.}~\bibnamefont {Nakamura}}, \bibinfo {author} {\bibfnamefont {J.}~\bibnamefont {Daniels}}, \bibinfo {author} {\bibfnamefont {C.}~\bibnamefont {Schroeder}}, \emph {et~al.},\ }\href@noop {} {\bibfield  {journal} {\bibinfo  {journal} {Physical review letters}\ }\textbf {\bibinfo {volume} {115}},\ \bibinfo {pages} {184802} (\bibinfo {year} {2015})}\BibitemShut {NoStop}%
\bibitem [{\citenamefont {Pompili}\ \emph {et~al.}(2018)\citenamefont {Pompili}, \citenamefont {Anania}, \citenamefont {Bellaveglia}, \citenamefont {Biagioni}, \citenamefont {Bini}, \citenamefont {Bisesto}, \citenamefont {Brentegani}, \citenamefont {Cardelli}, \citenamefont {Castorina}, \citenamefont {Chiadroni} \emph {et~al.}}]{pompili2018focusing}%
  \BibitemOpen
  \bibfield  {author} {\bibinfo {author} {\bibfnamefont {R.}~\bibnamefont {Pompili}}, \bibinfo {author} {\bibfnamefont {M.}~\bibnamefont {Anania}}, \bibinfo {author} {\bibfnamefont {M.}~\bibnamefont {Bellaveglia}}, \bibinfo {author} {\bibfnamefont {A.}~\bibnamefont {Biagioni}}, \bibinfo {author} {\bibfnamefont {S.}~\bibnamefont {Bini}}, \bibinfo {author} {\bibfnamefont {F.}~\bibnamefont {Bisesto}}, \bibinfo {author} {\bibfnamefont {E.}~\bibnamefont {Brentegani}}, \bibinfo {author} {\bibfnamefont {F.}~\bibnamefont {Cardelli}}, \bibinfo {author} {\bibfnamefont {G.}~\bibnamefont {Castorina}}, \bibinfo {author} {\bibfnamefont {E.}~\bibnamefont {Chiadroni}}, \emph {et~al.},\ }\href@noop {} {\bibfield  {journal} {\bibinfo  {journal} {Physical review letters}\ }\textbf {\bibinfo {volume} {121}},\ \bibinfo {pages} {174801} (\bibinfo {year} {2018})}\BibitemShut {NoStop}%
\bibitem [{\citenamefont {Lindstr{\o}m}\ \emph {et~al.}(2025)\citenamefont {Lindstr{\o}m}, \citenamefont {Adli}, \citenamefont {Anderson}, \citenamefont {Drobniak}, \citenamefont {Kalvik}, \citenamefont {Pe{\~n}a},\ and\ \citenamefont {Sjobak}}]{lindstrom2025sparta}%
  \BibitemOpen
  \bibfield  {author} {\bibinfo {author} {\bibfnamefont {C.}~\bibnamefont {Lindstr{\o}m}}, \bibinfo {author} {\bibfnamefont {E.}~\bibnamefont {Adli}}, \bibinfo {author} {\bibfnamefont {H.}~\bibnamefont {Anderson}}, \bibinfo {author} {\bibfnamefont {P.}~\bibnamefont {Drobniak}}, \bibinfo {author} {\bibfnamefont {D.}~\bibnamefont {Kalvik}}, \bibinfo {author} {\bibfnamefont {F.}~\bibnamefont {Pe{\~n}a}},\ and\ \bibinfo {author} {\bibfnamefont {K.}~\bibnamefont {Sjobak}},\ }\href@noop {} {\bibfield  {journal} {\bibinfo  {journal} {arXiv preprint arXiv:2505.14493}\ } (\bibinfo {year} {2025})}\BibitemShut {NoStop}%
\bibitem [{\citenamefont {Drobniak}\ \emph {et~al.}(2025)\citenamefont {Drobniak}, \citenamefont {Adli}, \citenamefont {Anderson}, \citenamefont {Dyson}, \citenamefont {Mewes}, \citenamefont {Sjobak}, \citenamefont {Th{\'e}venet},\ and\ \citenamefont {Lindstr{\o}m}}]{drobniak2025development}%
  \BibitemOpen
  \bibfield  {author} {\bibinfo {author} {\bibfnamefont {P.}~\bibnamefont {Drobniak}}, \bibinfo {author} {\bibfnamefont {E.}~\bibnamefont {Adli}}, \bibinfo {author} {\bibfnamefont {H.~B.}\ \bibnamefont {Anderson}}, \bibinfo {author} {\bibfnamefont {A.}~\bibnamefont {Dyson}}, \bibinfo {author} {\bibfnamefont {S.}~\bibnamefont {Mewes}}, \bibinfo {author} {\bibfnamefont {K.}~\bibnamefont {Sjobak}}, \bibinfo {author} {\bibfnamefont {M.}~\bibnamefont {Th{\'e}venet}},\ and\ \bibinfo {author} {\bibfnamefont {C.}~\bibnamefont {Lindstr{\o}m}},\ }\href@noop {} {\bibfield  {journal} {\bibinfo  {journal} {Nuclear Instruments and Methods in Physics Research Section A: Accelerators, Spectrometers, Detectors and Associated Equipment}\ ,\ \bibinfo {pages} {170223}} (\bibinfo {year} {2025})}\BibitemShut {NoStop}%
\bibitem [{\citenamefont {Rossi}\ \emph {et~al.}(2020)\citenamefont {Rossi}, \citenamefont {Petrillo}, \citenamefont {Bacci}, \citenamefont {Chiadroni}, \citenamefont {Cianchi}, \citenamefont {Ferrario}, \citenamefont {Giribono}, \citenamefont {Conti}, \citenamefont {Serafini},\ and\ \citenamefont {Vaccarezza}}]{rossi2020angstrom}%
  \BibitemOpen
  \bibfield  {author} {\bibinfo {author} {\bibfnamefont {A.~R.}\ \bibnamefont {Rossi}}, \bibinfo {author} {\bibfnamefont {V.}~\bibnamefont {Petrillo}}, \bibinfo {author} {\bibfnamefont {A.}~\bibnamefont {Bacci}}, \bibinfo {author} {\bibfnamefont {E.}~\bibnamefont {Chiadroni}}, \bibinfo {author} {\bibfnamefont {A.}~\bibnamefont {Cianchi}}, \bibinfo {author} {\bibfnamefont {M.}~\bibnamefont {Ferrario}}, \bibinfo {author} {\bibfnamefont {A.}~\bibnamefont {Giribono}}, \bibinfo {author} {\bibfnamefont {M.~R.}\ \bibnamefont {Conti}}, \bibinfo {author} {\bibfnamefont {L.}~\bibnamefont {Serafini}},\ and\ \bibinfo {author} {\bibfnamefont {C.}~\bibnamefont {Vaccarezza}},\ }in\ \href@noop {} {\emph {\bibinfo {booktitle} {Journal of Physics: Conference Series}}},\ Vol.\ \bibinfo {volume} {1596}\ (\bibinfo {organization} {IOP Publishing},\ \bibinfo {year} {2020})\ p.\ \bibinfo {pages} {012004}\BibitemShut {NoStop}%
\bibitem [{\citenamefont {Pompili}\ \emph {et~al.}(2024)\citenamefont {Pompili}, \citenamefont {Anania}, \citenamefont {Biagioni}, \citenamefont {Carillo}, \citenamefont {Chiadroni}, \citenamefont {Cianchi}, \citenamefont {Costa}, \citenamefont {Curcio}, \citenamefont {Crincoli}, \citenamefont {Del~Dotto} \emph {et~al.}}]{pompili2024guiding}%
  \BibitemOpen
  \bibfield  {author} {\bibinfo {author} {\bibfnamefont {R.}~\bibnamefont {Pompili}}, \bibinfo {author} {\bibfnamefont {M.}~\bibnamefont {Anania}}, \bibinfo {author} {\bibfnamefont {A.}~\bibnamefont {Biagioni}}, \bibinfo {author} {\bibfnamefont {M.}~\bibnamefont {Carillo}}, \bibinfo {author} {\bibfnamefont {E.}~\bibnamefont {Chiadroni}}, \bibinfo {author} {\bibfnamefont {A.}~\bibnamefont {Cianchi}}, \bibinfo {author} {\bibfnamefont {G.}~\bibnamefont {Costa}}, \bibinfo {author} {\bibfnamefont {A.}~\bibnamefont {Curcio}}, \bibinfo {author} {\bibfnamefont {L.}~\bibnamefont {Crincoli}}, \bibinfo {author} {\bibfnamefont {A.}~\bibnamefont {Del~Dotto}}, \emph {et~al.},\ }\href@noop {} {\bibfield  {journal} {\bibinfo  {journal} {Physical Review Letters}\ }\textbf {\bibinfo {volume} {132}},\ \bibinfo {pages} {215001} (\bibinfo {year} {2024})}\BibitemShut {NoStop}%
\bibitem [{\citenamefont {Frazzitta}\ \emph {et~al.}(2024)\citenamefont {Frazzitta}, \citenamefont {Pompili},\ and\ \citenamefont {Rossi}}]{frazzitta2024theory}%
  \BibitemOpen
  \bibfield  {author} {\bibinfo {author} {\bibfnamefont {A.}~\bibnamefont {Frazzitta}}, \bibinfo {author} {\bibfnamefont {R.}~\bibnamefont {Pompili}},\ and\ \bibinfo {author} {\bibfnamefont {A.}~\bibnamefont {Rossi}},\ }\href@noop {} {\bibfield  {journal} {\bibinfo  {journal} {Physical Review Accelerators and Beams}\ }\textbf {\bibinfo {volume} {27}},\ \bibinfo {pages} {091301} (\bibinfo {year} {2024})}\BibitemShut {NoStop}%
\bibitem [{\citenamefont {Banerjee}\ \emph {et~al.}(2021)\citenamefont {Banerjee}, \citenamefont {Bernhard}, \citenamefont {Brugger}, \citenamefont {Charitonidis}, \citenamefont {D’Alessandro}, \citenamefont {Gatignon}, \citenamefont {Gerbershagen}, \citenamefont {Montbarbon}, \citenamefont {Mussolini}, \citenamefont {Parozzi}, \citenamefont {Rae},\ and\ \citenamefont {Veit}}]{IPAC-M2}%
  \BibitemOpen
  \bibfield  {author} {\bibinfo {author} {\bibfnamefont {D.}~\bibnamefont {Banerjee}}, \bibinfo {author} {\bibfnamefont {J.}~\bibnamefont {Bernhard}}, \bibinfo {author} {\bibfnamefont {M.}~\bibnamefont {Brugger}}, \bibinfo {author} {\bibfnamefont {N.}~\bibnamefont {Charitonidis}}, \bibinfo {author} {\bibfnamefont {G.}~\bibnamefont {D’Alessandro}}, \bibinfo {author} {\bibfnamefont {L.}~\bibnamefont {Gatignon}}, \bibinfo {author} {\bibfnamefont {A.}~\bibnamefont {Gerbershagen}}, \bibinfo {author} {\bibfnamefont {E.}~\bibnamefont {Montbarbon}}, \bibinfo {author} {\bibfnamefont {C.}~\bibnamefont {Mussolini}}, \bibinfo {author} {\bibfnamefont {E.}~\bibnamefont {Parozzi}}, \bibinfo {author} {\bibfnamefont {B.}~\bibnamefont {Rae}},\ and\ \bibinfo {author} {\bibfnamefont {B.}~\bibnamefont {Veit}},\ }in\ \href {https://doi.org/10.18429/JACoW-IPAC2021-THPAB143} {\emph {\bibinfo {booktitle} {Proc. IPAC'21}}},\ \bibinfo {series and number} {\bibinfo {series} {International Particle Accelerator Conference}\ No.~\bibinfo
  {number} {12}}\ (\bibinfo  {publisher} {JACoW Publishing, Geneva, Switzerland},\ \bibinfo {year} {2021})\ pp.\ \bibinfo {pages} {4041--4044}\BibitemShut {NoStop}%
\bibitem [{\citenamefont {Floettmann}(1997)}]{floettmann2011astra}%
  \BibitemOpen
  \bibfield  {author} {\bibinfo {author} {\bibfnamefont {K.}~\bibnamefont {Floettmann}},\ }\href {http://www.desy.de/~mpyflo/} {\bibinfo {title} {Astra: A space charge tracking algorithm}} (\bibinfo {year} {1997})\BibitemShut {NoStop}%
\bibitem [{\citenamefont {Giuliano}\ \emph {et~al.}(2025)\citenamefont {Giuliano}, \citenamefont {Alesini}, \citenamefont {Cardelli}, \citenamefont {Carillo}, \citenamefont {Chiadroni}, \citenamefont {Coppola}, \citenamefont {Cuttone}, \citenamefont {Curcio}, \citenamefont {De~Gregorio}, \citenamefont {Di~Raddo} \emph {et~al.}}]{giuliano2025compact}%
  \BibitemOpen
  \bibfield  {author} {\bibinfo {author} {\bibfnamefont {L.}~\bibnamefont {Giuliano}}, \bibinfo {author} {\bibfnamefont {D.}~\bibnamefont {Alesini}}, \bibinfo {author} {\bibfnamefont {F.}~\bibnamefont {Cardelli}}, \bibinfo {author} {\bibfnamefont {M.}~\bibnamefont {Carillo}}, \bibinfo {author} {\bibfnamefont {E.}~\bibnamefont {Chiadroni}}, \bibinfo {author} {\bibfnamefont {M.}~\bibnamefont {Coppola}}, \bibinfo {author} {\bibfnamefont {G.}~\bibnamefont {Cuttone}}, \bibinfo {author} {\bibfnamefont {A.}~\bibnamefont {Curcio}}, \bibinfo {author} {\bibfnamefont {A.}~\bibnamefont {De~Gregorio}}, \bibinfo {author} {\bibfnamefont {R.}~\bibnamefont {Di~Raddo}}, \emph {et~al.},\ }\href@noop {} {\bibfield  {journal} {\bibinfo  {journal} {Frontiers in Oncology}\ }\textbf {\bibinfo {volume} {15}},\ \bibinfo {pages} {1516576} (\bibinfo {year} {2025})}\BibitemShut {NoStop}%
\bibitem [{\citenamefont {Valente}\ \emph {et~al.}(2006)\citenamefont {Valente}, \citenamefont {Buonomo},\ and\ \citenamefont {Mazzitelli}}]{valente2006diagnostics}%
  \BibitemOpen
  \bibfield  {author} {\bibinfo {author} {\bibfnamefont {P.}~\bibnamefont {Valente}}, \bibinfo {author} {\bibfnamefont {B.}~\bibnamefont {Buonomo}},\ and\ \bibinfo {author} {\bibfnamefont {G.}~\bibnamefont {Mazzitelli}},\ }\href@noop {} {\bibfield  {journal} {\bibinfo  {journal} {Nuclear Physics B-Proceedings Supplements}\ }\textbf {\bibinfo {volume} {150}},\ \bibinfo {pages} {362} (\bibinfo {year} {2006})}\BibitemShut {NoStop}%
\bibitem [{\citenamefont {Korotchenko}\ \emph {et~al.}(2020)\citenamefont {Korotchenko}, \citenamefont {Rozhkova},\ and\ \citenamefont {Dabagov}}]{Korotchenko2020}%
  \BibitemOpen
  \bibfield  {author} {\bibinfo {author} {\bibfnamefont {K.~B.}\ \bibnamefont {Korotchenko}}, \bibinfo {author} {\bibfnamefont {E.~I.}\ \bibnamefont {Rozhkova}},\ and\ \bibinfo {author} {\bibfnamefont {S.~B.}\ \bibnamefont {Dabagov}},\ }\href {https://doi.org/10.1140/epjc/s10052-020-08492-9} {\bibfield  {journal} {\bibinfo  {journal} {European Physical Journal C}\ }\textbf {\bibinfo {volume} {80}},\ \bibinfo {pages} {927} (\bibinfo {year} {2020})},\ \bibinfo {note} {open Access}\BibitemShut {NoStop}%
\bibitem [{\citenamefont {Hrouda}(1982)}]{hrouda1982magnetic}%
  \BibitemOpen
  \bibfield  {author} {\bibinfo {author} {\bibfnamefont {F.}~\bibnamefont {Hrouda}},\ }\href {https://doi.org/10.1007/BF01450244} {\bibfield  {journal} {\bibinfo  {journal} {Geophysical Surveys}\ }\textbf {\bibinfo {volume} {5}},\ \bibinfo {pages} {37} (\bibinfo {year} {1982})}\BibitemShut {NoStop}%
\bibitem [{\citenamefont {ichiro Otofuji}\ \emph {et~al.}(2000)\citenamefont {ichiro Otofuji}, \citenamefont {Uno}, \citenamefont {Higashi}, \citenamefont {Ichikawa}, \citenamefont {Ueno}, \citenamefont {Mishima},\ and\ \citenamefont {Matsuda}}]{OTOFUJI2000271}%
  \BibitemOpen
  \bibfield  {author} {\bibinfo {author} {\bibfnamefont {Y.}~\bibnamefont {ichiro Otofuji}}, \bibinfo {author} {\bibfnamefont {K.}~\bibnamefont {Uno}}, \bibinfo {author} {\bibfnamefont {T.}~\bibnamefont {Higashi}}, \bibinfo {author} {\bibfnamefont {T.}~\bibnamefont {Ichikawa}}, \bibinfo {author} {\bibfnamefont {T.}~\bibnamefont {Ueno}}, \bibinfo {author} {\bibfnamefont {T.}~\bibnamefont {Mishima}},\ and\ \bibinfo {author} {\bibfnamefont {T.}~\bibnamefont {Matsuda}},\ }\href {https://doi.org/https://doi.org/10.1016/S0012-821X(00)00169-2} {\bibfield  {journal} {\bibinfo  {journal} {Earth and Planetary Science Letters}\ }\textbf {\bibinfo {volume} {180}},\ \bibinfo {pages} {271} (\bibinfo {year} {2000})}\BibitemShut {NoStop}%
\end{thebibliography}

%

\end{document}